
\magnification = \magstep1

\def\item{\vskip1.3pt\hang\textindent\rm}


\tolerance=300
\pretolerance=200
\hfuzz=1pt
\vfuzz=1pt

\hoffset 0cm     
\voffset=0in
\hsize=5.8 true in
\vsize=9.5 true in

\parindent=35pt
\mathsurround=1pt
\parskip=1pt plus .25pt minus .25pt
\normallineskiplimit=.99pt

\mathchardef\emptyset="001F 


%

\def\Aut{\mathop{\rm Aut}\nolimits}

\def\1{{\bf 1}} \def\0{\bf 0}

\def\({\bigl(}  \def\){\bigr)}
\def\<{\mathopen{\langle}}\def\>{\mathclose{\rangle}}

\def\({\bigl(}  \def\){\bigr)}
\def\<{\mathopen{\langle}}\def\>{\mathclose{\rangle}}
\def\Z{{\mathchoice{{\hbox{$\rm Z\hskip 0.26em\llap{\rm Z}$}}}%
{{\hbox{$\rm Z\hskip 0.26em\llap{\rm Z}$}}}%
{{\hbox{$\scriptstyle\rm Z\hskip 0.31em\llap{$\scriptstyle\rm Z$}$}}}{{%
\hbox{$\scriptscriptstyle\rm Z$\hskip0.18em\llap{$\scriptscriptstyle\rm
Z$}}}}}}

\def\N{{\mathchoice{\hbox{$\rm I\hskip-0.14em N$}}%
{\hbox{$\rm I\hskip-0.14em N$}}%
{\hbox{$\scriptstyle\rm I\hskip-0.14em N$}}%
{\hbox{$\scriptscriptstyle\rm I\hskip-0.10em N$}}}}

\def\R{{\mathchoice{\hbox{$\rm I\hskip-0.14em R$}}%
{\hbox{$\rm I\hskip-0.14em R$}}%
{\hbox{$\scriptstyle\rm I\hskip-0.14em R$}}%
{\hbox{$\scriptscriptstyle\rm I\hskip-0.10em R$}}}}

\def\C{{\mathchoice%
{\hbox{$\rm C\hskip-0.47em\hbox{%
\vrule height 0.58em width 0.06em depth-0.035em}$}\;}%
{\hbox{$\rm C\hskip-0.47em\hbox{%
\vrule height 0.58em width 0.06em depth-0.035em}$}\;}%
{\hbox{$\scriptstyle\rm C\hskip-0.46em\hbox{%
$\scriptstyle\vrule height 0.365em width 0.05em depth-0.025em$}$}\;}%
{\hbox{$\scriptscriptstyle\rm C\hskip-0.41em\hbox{
$\scriptscriptstyle\vrule height 0.285em width 0.04em depth-0.018em$}$}\;}}}

\def\lastname{Author, no initials????}     \def\date{}
\def\rightheadline{\nin\eightrm\date \hfil\smc\lastname\hfil\tenbf\folio}
\def\leftheadline{\tenbf\folio\hfil\smc\lastname\hfil\eightrm\date}
\headline={\ifodd\pageno\rightheadline\else\leftheadline\fi}

\def\sectionheadline #1{\bigbreak\vskip-\lastskip\indent\vskip1.5cm
                       \centerline{\bf #1}\nobreak\medskip\nobreak}

\def\.{{\cdot}}
\def\|{\Vert}

\def\msk{\medskip}
\def\bsk{\bigskip}
\def\giantskip{\vskip2\bigskipamount}

\def\bbr{\bigbreak}
\def\giantbreak{\par \ifdim\lastskip<2\bigskipamount \removelastskip
         \penalty-400 \giantskip\fi}

\def\nin{\noindent}
\def\cen{\centerline}
\def\pagebreak{\vskip 0pt plus 0.0001fil\break}
\def\linebreak{\break}

\def\epsilon{\varepsilon}
\def\phi{\varphi}

\def\tilde{\widetilde}

\def\hat{\widehat}

\font\eightrm=cmr8
\font\sixrm=cmr6


\font\bfone=cmbx10 scaled\magstep1 
\font\bftwo=cmbx10 scaled\magstep2 
\font\smc=cmcsc10

\def\nin{\noindent}

\def\Proposition #1. {\bigbreak\vskip-\parskip\noindent{\bf #1. \quad
 Proposition.}\quad\it}

\def\Theorem #1. {\bigbreak\vskip-\parskip\noindent{\bf #1. \quad
 Theorem.}\quad\it}
\def\Corollary #1. {\bbr\vskip-\parskip\nin{\bf #1. \quad Corollary.}
\quad\it}
\def\Lemma #1. {\bigbreak\vskip-\parskip\noindent{\bf #1. \quad
 Lemma.}\quad\it}

\def\Definition #1. {\rm\bigbreak\vskip-\parskip\noindent{\bf #1. \quad
 Definition.}\quad}

\def\Remark #1. {\rm\bigbreak\vskip-\parskip\noindent{\bf #1. \quad
 Remark.}\quad}

\def\Remarks #1. {\rm\bigbreak\vskip-\parskip\noindent{\bf #1. \quad
 Remarks.}\quad}

\def\Exercise #1. {\rm\bigbreak\vskip-\parskip\noindent{\bf #1. \quad
Exercise.}\quad}

\def\Exercises #1. {\rm\bigbreak\vskip-\parskip\noindent{\bf #1. \quad
Exercises.}\quad}

\def\Example #1. {\rm\bigbreak\vskip-\parskip\noindent{\bf #1. \quad
Example.}\quad}

\def\Examples #1. {\rm\bigbreak\vskip-\parskip\noindent{\bf #1. \quad
Examples.}\quad}

\def\Proof#1.{\rm\par\ifdim\lastskip<\bigskipamount\removelastskip\fi\smallskip
            \noindent {\bf Proof.}\quad}

\nopagenumbers

\def\lastname{}
\def\dom{\hbox{\rm dom$\,$}}
\def\no#1. {\removelastskip\bsk\noindent {\bf#1. }}
\def\eop{\par\line{\hfill \bf e.~o.~p.}\msk\rm}
\newcount\Lemmanr
\def\Lemma{\advance \Lemmanr by1\removelastskip\msk\noindent {\bf Lemma
\the\Lemmanr. } \it}
\newcount\Definr
\def\Definition{\advance \Definr by1\removelastskip\msk\noindent {\bf
Definition\the\Definr. }}
\let\eitem\item
\def\item#1{\eitem{\rm#1}}
\vbox{\vskip 3cm}

\def\span{{\rm span\ }}
\def\map{{\rm map\ }}

\def\[#1 {\smallskip\hang\textindent{\rm[#1]} } 

\cen{\bftwo Effective Quantum Observables}
\bsk
\cen{\bfone M.~Baaz$^1$, N.~Brunner$^2$ and K.~Svozil$^{3}$}
\footnote{}{\hskip-\parindent\eightrm\nin $^1$
Abt.~Theoretische Informatik, TU Wien, Wiedner
Hauptstra\ss e 8--10, A--1040 Wien}
\footnote{}{\hskip-\parindent\eightrm\nin $^2$  Inst. Math., U. Bodenkultur,
Gregor Mendel--Str.33, A--1180 Wien}
\footnote{}{\hskip-\parindent\eightrm \nin$^3$
Inst.~Theoretische Physik, TU Wien, Wiedner
Hauptstra\ss e 8--10, A--1040 Wien}
\msk\msk

{\eightrm
\nin ABSTRACT: Thought experiments about the physical nature of set theoretical
counterexamples to the axiom of choice motivate the investigation of peculiar
constructions, e.g. an infinite dimensional Hilbert space with a modular
quantum logic. Applying a concept due to B{\sixrm ENIOFF}, we identify the
intrinsically effective Hamiltonians with those observables of quantum theory
which may coexist with a failure of the axiom of choice. Here a self
adjoint operator is intrinsically effective, iff the Schr\"odinger equation
of its generated semigroup is soluble by means of eigenfunction series
expansions.}
\msk\msk

\sectionheadline{\bf Introduction and Summary}
\msk\msk

\no {0.1. The problem}.
The formalism of quantum theory is known to depend in an essential
way on infinite structures, whence one might wonder about the effects of
infinity.
A proper analysis of this problem seems to require the reconstruction of
quantum theory within some system of constructive mathematics.
\msk

{\smc Hellmann}
[He93] has questioned the feasibility of the standard approach towards
constructive quantum theory by means of
computability theory. Instead we
investigate effective quantum theory which rest on a very mild condition of
constructivity. A concept is {\it effective}
in the sense of {\smc Sierpinski}, if
it does not require the axiom of choice. Nevertheless, even the very notion of
a self adjoint operator as an observable of quantum theory may  become
meaningless
in this context. This motivates the question, if there is an alternative
definition for observables which does not depend on the
axiom of choice (AC), and if so, what is its relation to the usual one.
\msk\msk

\no{0.2. A solution}. Our point of departure is the {\smc Kochen} and {\smc
Specker} [Ko67] definition of a quantum theoretical observable on a finite
dimensional Hilbert space. We apply general requirements about the notion of
measurement in empirical structures in order to extend this definition to
infinite dimensional spaces. The definition of these finitary observables does
not depend on $AC$. Not surprisingly then, finitary observables do not
correspond to self adjoint operators but to a different class of finitary
mappings on Hilbert space. There is, however, a class of quantum like finitary
mappings which have the same functional calculus as the quantum theoretical
observables.
\msk

While in the presence of $AC$, these quantum like observables
reduce essentially to the finite dimensional observables of quantum theory,
there are
permutation models of $ZFA$ set theory minus $AC$ which
admit nontrivial quantum
like observables. $ZFA$ is a variant of $ZF$ set theory which
permits a set $A$ of atoms
(objects without elements; c.f. {\smc Jech} [Je73]).
Investigations into the physical nature of {\smc Russell}'s
socks provide even a physical motivation for these models. Moreover, the
quantum like observables in such models may be extended to self adjoint
operators of the surrounding world $ZFC\ (=ZF+AC)$ in a way which respects the
functional calculus. Our proposed constructive variant of quantum theory
is the $ZFC$
theory of these intrinsically effective self adjoint operators.
\msk

Implicitly, intrinsically effective
Hamiltonians have been considered in thought experiments of a
quantum philosophical nature. For example in 1987, in the context of
speculations about quantum chaos,
A. {\smc Peres} (c.f. [Pe93]) has determined as non--chaotic those
Hamiltonians,  whose Schr\"odinger equations may be solved by means
of eigenfunction series expansions. The latter property characterizes
intrinsic effectivity (lemma 16.)
\msk\msk

\no {0.3. Acknowledgements}. Preliminary versions of our paper have been
presented in June 1992 at a conference of the
DVMLGdM at M\"unster and in
March 1993 at a meeting of the contact group in mathematical logic at Louvaine
de la Neufe. The
discussions at these meetings have helped to improve the presentation of the
present paper.
\msk

Thanks for her help with the preparation of the
manuscript are due to Mrs. {\smc Holy} (U. BoKu.).
The authors gratefully remember helpful comments
by professors {\smc Ruppert} (U. BoKu.),
{\smc Hejtmanek} and {\smc Schachermayer} (University of Vienna).
\msk\msk

\sectionheadline{\bf 1. A general notion of observables}
\msk\msk

\no {1.1. Observations}.
A measurement is an assignment of mathematical constructs to the
elements of an empirical structure. Any particular measurement may be
represented by the data which it produces. Thus if $X$ is the empirical
structure in question, $Y$ is the set of intended values and $E\subseteq X$ is
the set of states which are relevant for the measurement situation, then the
set $f = \{(x,fx):x\in E\}$ of all collected data forms a function $f: E\to
Y$ which describes the measurement in an extensional way.
\msk

We let ${\cal F}(X,Y)$ be the set of all partial functions from $X$ to $Y$ and
represent all possible measurements by a set ${\cal A}\subseteq{\cal
F}(X,Y)$ of allowable  functions.
\msk\msk

\no {1.2. Information}. The measurements are quasiordered according to their
information
content by a reflexive and transitive relation $\le$ on ${\cal A}$. If
$f\subseteq g\in{\cal A}$ (set inclusion), then we require $f\in{\cal A}$ and
$f\le g$, since $f$ can be obtained from $g$ by forgetting data. The
quasiordering reflects the predictions which can be drawn from an empirical
theory.
\msk\msk

\no {1.3. Knowledge}.
An observable ${\cal O}\subseteq{\cal A}$ shall be a nonempty
collection of measurements which represents a state of knowledge corresponding
to measurements from the same class of procedures. The following conditions
seem natural: (i) If $f\le g\in{\cal O}$, then $f\in{\cal O}$, since $f$
contains less information. (ii) If $f\in{\cal
O}$ and $g\in{\cal O}$, then for some $h\in{\cal O}$ both $f\le h$ and
$g\le h$; $h$ corresponds to an increased amount of information by the
repeated application of the same experimental scheme. Thus ${\cal O}$ is an
order ideal in ${\cal A}$.
\msk

The ideal ${\cal O} = \{\emptyset\}$ corresponds to ``no empirical evidence".
Observables are idealizations which represent maximal knowledge. We
arrive at the following definition.
\msk

{\it An observable ${\cal O}$ for an empirical structure
$({\cal A},\le$) is a maximal order ideal in ${\cal A}$.}
\msk\msk

\no {1.4. Functions}.
An important feature of observables is the existence of an functional
calculus resembling that on $Y$.
\msk

If $F\colon Y^n\to Y$ and $f_1,\cdots f_n\in\hbox{$Y^E$}$ are functions, then
$F{\scriptstyle\circ}(f_1,\cdots f_n)\in\hbox{$Y^E$}$ is defined as
$F{\scriptstyle\circ}(f_1,\cdots f_n)(a) = F(f_1(a),\cdots f_n(a))$.
For $F$ we assume, that
$F{\scriptstyle\circ}(f_1,\cdots f_n)\in{\cal A}$
for all $f_i\in{\cal A}$ with the same domain
and that $F{\scriptstyle\circ}(f_1,\cdots
f_n)\le F{\scriptstyle\circ}(g_1,\cdots g_n)$
for all $f_i\in{\cal A}\cap\hbox{$Y^E$}$,
$g_i\in{\cal
A}\cap\hbox{$Y^G$}$ such that $f_i\le g_i$ for all $i$.
\msk

Ideals ${\cal O}_1,\cdots{\cal O}_n$ are {\it commeasurable}, if for each
sequence
$f_i\in{\cal O}_i$ of functions there exist $g_i\in{\cal O}_i$ such that
$g_i\ge f_i$ and all functions $g_i$ have the same domain. With $F$ we
associate a function $\hat F$ which is defined on commeasurable ideals ${\cal
O}_1,\cdots{\cal O}_n: \hat F({\cal O}_1,\cdots{\cal O}_n) = \{f\in{\cal A}:
f\le F{\scriptstyle\circ}(g_1,\cdots g_n)$
for some sequence $g_i\in{\cal O}_i$ of functions
with the same domain$\}$. This is an ideal and ${\cal O}_1,\cdots{\cal O}_n,
\hat F({\cal O}_1,\cdots{\cal O}_n)$ are commeasurable. However, even if all
${\cal O}_i$ are observables, $\hat F({\cal O}_1,\cdots{\cal O}_n)$ may fail
to be maximal. \msk

A function $F\colon Y^n\to Y$ is {\it admissible}, if in addition to the
requirements for the definition of $\hat F$, $\hat F$ maps commeasurable
observables to observables. In section 1.5 there is a counterexample which
shows, that the set of admissible functions in general is not closed under
composition. For observables of quantum theory, commeasurability coincides
with commensurability (c.f. [Ko67]) and the functional calculus is the usual
one.
\msk\msk

\no {1.5. Counterexample}. We illustrate our terminology by a simple notion of
measurement, where names $y\in Y$ are attributed to the objects $x\in X$ and
renaming or the identification of names decreases
the information content. To this end we
consider the structure $({\cal A},\le)$, where
${\cal A} = \{f\in{\cal F}(X,Y)\colon\ \dom f$ is finite$\}$
and $f\le g$, iff $\dom f\subseteq\dom g$
and there is a function $h\colon Y\to Y$ such that
$f=h{\scriptstyle\circ}g\mid \dom f$ [that is, $\forall x\in \dom f\colon
f(x)=h\(g(x)\)$.]
\msk

\Lemma  {\it In $ZFA$, if $Y$ is
infinite, then the set $({\cal I},\subseteq)$ of all order ideals of $({\cal
A},\le)$ is isomorphic to the complete lattice $(\Pi,\succeq)$ of all
partitions ${\cal P}$ of $X^+ = X\cup\{a\}$, $a\notin X$ an additional
distinguished object, where ${\cal P}\preceq{\cal Q}$, iff ${\cal P}$ is a
refinement of ${\cal Q}$.}
\msk

\Proof. If ${\cal P}$ is a partition of $X^+$, we define $j({\cal
P})\subseteq{\cal A}$ as $j({\cal P}) = \{f\in{\cal A}\colon\dom f\cap P =
\emptyset$, if $a\in P\in{\cal P}$, and $fx = fy$, if $\{x,y\}\subseteq P$ for
some $P\in{\cal P}\}$. Obviously, $\emptyset\in j{\cal P}$ and $f\le g\in
j{\cal (P)}$ implies $f\in j{\cal (P)}$. If $f$ and $g$ are in $j{\cal (P)}$
and $E = \dom f\cup\dom g$ intersects exactly $n$ elements $P_1,\cdots P_n$ of
${\cal P}$, we let $y_1,\cdots y_n$ be arbitrary $n$ elements of $Y$ and
define $h\colon E\to Y$ as $h(x) = y_i$, if $x\in P_i$. Then $h\in j{\cal
(P)}$ and $h\ge f$, $h\ge g$, whence $j{\cal (P)}$ is an order ideal.
\msk

If ${\cal O}\subseteq{\cal A}$ is an order ideal, we define the following
relation $\sim$ on $X^+\colon \dom {\cal O} = \cup\{\dom f\colon f\in{\cal
O}\}$ and $x\sim y$, iff $\{x,y\}\cap\dom{\cal O} = \emptyset$ or
$\{x,y\}\subseteq\dom{\cal O}$ and $fx = fy$ for all $f\in{\cal O}$ such that
$\{x,y\}\subseteq\dom f$. Note that, if $E\subseteq\dom{\cal O}$ is finite,
then there is a function $f_E\in{\cal O}$ such that $E\subseteq\dom f_E$ (if
$x\in\dom f_x$ and $f_x\in{\cal O}$ for $x\in E$, then let $f_E\in{\cal O}$ be
an upper bound of these finitely many $f_x$). This defines an equivalence
relation, for if $x\sim y$, $y\sim z$ and $\{x,y,z\}\subseteq\dom{\cal O}$, but
not
$x\sim z$, then for some $f\in{\cal O}\ \dom f\supseteq\{x,y,z\}$ and for some
$g\in{\cal O}\  gx\not=gz$. If $h\in{\cal O}$ and $h\ge f,g$, then $hx\not=hz$
and $\dom h\supseteq\{x,y,z\}$. It follows, that $hx\not=hy$ or $hy\not=hz$, a
contradiction, whence $\sim$ is transitive. We let $\pi{\cal O}$ be the
associated partition of $X^+$.
\msk

One easily verifies, that $\pi{\cal O}\preceq\pi{\cal O}'$, if ${\cal
O}\supseteq{\cal O}'$, and $j{\cal P}\subseteq j{\cal P}'$, if ${\cal
P}'\preceq{\cal P}$. Moreover $\pi j{\cal P} = {\cal P}$ for all partitions
${\cal P}$ and $j\pi{\cal O} = {\cal O}$ for all order ideals ${\cal O}$.
Concerning the latter identity, ${\cal O}\subseteq j\pi{\cal O}$ is obvious. If
conversely $f\colon E\to Y$ is in $j\pi{\cal O}$, then $E\subseteq\dom{\cal
O}$ and moreover $fx\not=fy$ implies $g_{xy}(x)\not=g_{xy}(y)$ for some
$g_{xy}\in{\cal O}$. We let $g\in{\cal O}$ be an upper bound of $f_E$ and all
$g_{xy}$. Then $g\ge f$ and therefore $f\in{\cal O}$.
\eop

Since ${\cal A}$ is an order ideal, it is the only observable.
As follows from lemma 1,
if $X$ and $Y$ are infinite, then ${\cal O}$ is an observable, iff for each
finite set of objects there exists a measurement in ${\cal O}$ which assigns
different names to these objects. Here the cardinality
of $Y$ needs not be larger than that of $X$. Therefore
a mapping $f:Y\to Y$, $Y$ infinite, is admissible
for the functional calculus, iff its range is
infinite. [If the range of $f$ is finite, then $\hat f{\cal A}$ is not an
ideal, while if the range of $f$ is infinite, then $\hat f{\cal A} = {\cal
A}$.] The family of these mappings is not closed under composition.
\msk

An order ideal ${\cal
O}$ is {\it scalable}, if there exists a function $f\colon X\to Y$
such that ${\cal O} = {\cal O}(f)$, where ${\cal O}(f)
= \{g\in{\cal A}\colon g\le f\mid E$ for some finite $E\subseteq X\}$. Since
$\pi{\cal O}(f) = \{\{a\}, f^{-1''}\{y\}\colon y\in f''X\}$, ${\cal A}$ is
scalable, iff there exists an
injective function $f\colon X\to Y$, a so called {\it cardinal scale}. This
notion is taken from mathematical psychology.
In $ZFA$, ${\cal O}(f)\supseteq{\cal O}(g)$, iff there is a function
$h\colon Y\to Y$ such that $g = h{\scriptstyle\circ} f$
(c.f. {\smc Hindman} and
{\smc Milnes} [Hi84] for related topological conditions.) In general $\hat
h{\cal O}(f)\neq {\cal O}(h{\scriptstyle\circ}f)$. [If $f(x)\neq f(y)$ and
$h{\scriptstyle\circ}f(x) = h{\scriptstyle\circ}f(y)$ and the range of $h$ is
infinite, then $\hat h{\cal O}(f) = {\cal O}(f)\neq {\cal
O}(h{\scriptstyle\circ}f)$.]
\msk

The {\it partition principle} is the assertion, that the injective and the
surjective orderings of cardinals coincide. It
is known to imply the wellorderable axiom
of choice ({\smc D. Pincus} in [Pe78], p. 587),
but it is still an open question, if it is equivalent to $AC$.
As follows from lemma 2 below, in $ZF$ set theory the partition principle
is equivalent to the
assertion, that for all structures $({\cal A},\le )$ with
infinite sets of names the
set of all scalable ideals is closed under bounded joins of order ideals.
\msk

\Lemma {\it In $ZF$, if $Y$ is infinite, then the join of a nonempty
family of scalable ideals which are bounded from above by a
scalable ideal is again scalable, iff the following holds: If
$\mu\le_*\kappa\le_*\vert X\vert$ and $\kappa\le\vert Y\vert$, then
$\mu\le\vert Y\vert$.}

\msk

\Proof. $\vert X\vert$ is the {\smc Scott} cardinality of $X$.
$\vert X\vert\le\vert Y\vert$,
if there exists an injective $f\colon X\to Y$ and $\vert X\vert\le_*\vert
Y\vert$, if there exists a surjective $f\colon Y\to X$.
\msk

The representable ideals are order isomorphic to
$(\Pi',\succeq),\Pi'$ the set
of all partitions ${\cal P'}$ of $X$ such that $\vert{\cal P'}\vert\le\vert
Y\vert$ ($\pi'({\cal O}) = \pi({\cal O})\setminus\{\{a\}\}$ defines the
isomorphism). Since for a partition ${\cal P}$ of $X\quad \vert{\cal
P}\vert\le_*\vert X\vert$ and for partitions ${\cal P}\preceq{\cal
Q}\quad\vert{\cal P}\vert\ge_*\vert{\cal Q}\vert$, the assumptions on the
cardinals imply the Dedekind-completeness of the partially ordered set $\Pi'$.
For if ${\cal L}\preceq{\cal P}$ for all ${\cal P}\in\Psi\subseteq\Pi'$ and
some ${\cal L}\in\Pi'$, then ${\cal L}\preceq\inf\Psi$ (formed in the set of
all partitions of $X$) and $\vert\inf\Psi\vert\le_*\vert{\cal
L}\vert\le_*\vert X\vert$, whence $\inf\Psi\in\Pi'$.
\msk

If conversely $\mu\le_*\kappa\le_*\vert X\vert$ and $\kappa\le\vert Y\vert$,
$\mu = \vert M\vert$ and $\kappa = \vert K\vert$, then a surjective mapping
$f\colon X\to K$ defines a partition ${\cal P} = \{f^{-1''}\{y\}\colon y\in
f''X\subseteq K\}\in\Pi'$ such that there is a surjective $g\colon{\cal P}\to
M$. We set ${\cal Q} = \{\cup g^{-1''}\{m\}\colon m\in M\}$, a partition of
$X$ of cardinality $\mu$. Since ${\cal Q} = \inf\Psi$ in the set of all
partitions of $X$, where $\Psi = \{\{\cup g^{-1''}\{m\}$, $\cup g^{-
1''}(M\setminus\{m\})\}\colon m\in M\}\subseteq\Pi'$, and because ${\cal
P}\preceq{\cal L}$ for all ${\cal L}\in\Psi$, ${\cal Q} = \inf\Psi\in\Pi'$, if
$\Pi'$ is closed under bounded infima. In this case $\mu = \vert{\cal
Q}\vert\le\vert Y\vert$.
\eop

There is a similar theory of preference structures. In the definition of
$({\cal A},\le )$ we let $(Y,<)$ be an infinite dense linear order without
endpoints and $f\le g$, iff $f = h{\scriptstyle\circ} g\mid \dom f$,
where $h:Y\to Y$ is
nondecreasing. Then as in lemma 1 the observables on $X$ correspond to the
linear orderings on $X$. Hence the {\it ordering principle} (each set is
orderable) is equivalent with the assertion, that each set admits
observables. Observables are commeasurable, iff their induced orderings
coincide.
\msk \msk

\sectionheadline{\bf 2. An effective version of quantum theory}
\msk\msk

\no {2.1. Finitary quantum mechanics}. We consider the finitary quantum
structure ${\cal A} =\{f\in{\cal F}(H,{\R})\colon\dom f$
a finite $ON$system$\}$ and $f\le g$, iff $\tilde f\subseteq\tilde g$, where
$\tilde f\colon\span \dom f\to H$ is given by
$$\tilde f(x) = \sum_{a\in\dom f}f(a)\cdot <x,a>\cdot a\,.$$
\msk

Here $H$ is a complex Hilbert space (completeness is defined in terms of
Cauchy sequences) of pure states with the scalar product $<\cdot ,\cdot >$.
The {\it finitary observables} on $H$ are
the observables for this structure.
\msk

If $H$ is finite dimensional, then this notion is equivalent with the
definition of the observables of quantum theory in {\smc Kochen} and {\smc
Specker} [Ko65]. In general, this is no longer true.
\msk\msk

\no {2.2. Functional calculus}. As has been pointed out by {\smc Kochen} and
{\smc Specker} [Ko67], in order to correspond to a quantum theoretical
observable, a finitary observable needs to obey the same functional calculus.
We now identify these quantum like observables with a class of linear
mappings. Symmetric linear mappings which satisfy the condition of the
following lemma are {\it finitary.}

\Lemma  {\it In $ZFA$ the set $({\cal I},\subseteq)$ of the ideals of $({\cal
A},\le)$ is isomorphic with the set $({\cal S},\subseteq)$ of all
symmetric linear operators $A\colon
\dom A\to H$, where $\dom A$ is the linear
span (not necessarily closed) of the set $EV(A)$ of all eigenvectors of
$A$.}
\msk

\Proof. If ${\cal O}\subseteq{\cal A}$ is an ideal, we set $\mu({\cal
O}) = \cup\{\tilde f : f\in{\cal O}\}$. Since ${\cal O}$ is directed,
$\mu({\cal O})$ is a function and if $E\subseteq\dom \mu({\cal O})$ is
finite, then $E\subseteq\dom \tilde f$ for some $f\in{\cal O}$, whence
$\mu({\cal O})$ is linear and symmetric. By definition $\dom\tilde f =
\span EV(\tilde f)$ and since $EV(\tilde f)\subseteq EV(\mu({\cal O}))$,
$\dom\mu({\cal O}) = \span EV(\mu({\cal O}))$, i.e. $\mu({\cal
O})\in{\cal S}$.
\msk

If $A\in{\cal S}$ and $E\subseteq EV(A)$ is a finite $ON$system, then we define
$A_E\colon E\to {\C}$ through $A_E(a)\cdot a = A(a), a\in E$. Since the
eigenvalues of symmetric matrices are real, $A_E\in{\cal A}$. We set $j(A) =
\{f\in {\cal A}\colon f\le A_E$ for some finite $ON$system $E\subseteq
EV(A)\}$ and observe, that $j(A)$ is an order ideal. For if $E$, $F$ are
finite $ON$systems in $EV(A)$, then $A_E, A_F\le A_G$, where $G\subseteq
EV(A\mid\span (E\cup F))$ is an $ON$system such that $\span G = \span (E\cup
F)$ which exists by diagonalization.
\msk

If $A\in{\cal S}$, then $\mu j A = \cup\{\tilde A_E\colon E\subseteq EV(A)$ a
finite $ON$system$\} = \cup\{A\mid\span E\colon E\subseteq EV(A)$ a finite set
$\} = A$. If ${\cal O}\in{\cal I}$, then $j\mu{\cal O} = \{f\colon
f\le\mu({\cal O})_E$ for some finite $ON$system $E\subseteq EV(\mu({\cal
O}))\}$. Since $\mu({\cal O})_E\le g$; whenever $g\in{\cal O}$
satisfies $\dom \tilde g\supseteq E$, $j\mu{\cal O}\subseteq {\cal O}$,
while $j\mu{\cal O}\supseteq{\cal O}$ is obvious.
\eop
\msk

If the finitary mapping $A$ satisfies $\dom A = H$, then $A$ corresponds to a
finitary observable. Observables of this kind (and the corresponding maps) are
{\it quantum like}.
\msk

As a motivation for this terminology, we note that one may
change the definition of a finitary observable, so
that in $ZFC$ the new class of observables includes the usual observables of
quantum theory. Let $({\cal A},\le)$ consist of
the symmetric linear maps, partially ordered by
inclusion. Then observables are a direct sum of a self--adjoint operator
and a finite number of unbounded operators $+ T$ or $- T$, where $T =
i(1 + S)(1 - S)^{-1}$, $S$ the unilateral shift on $\ell_2(\omega)$
({\smc von Neumann} [vN29]).
\msk

There is also an analogy with nonstandard analysis. We let $*$ be an
$\aleph_1$--saturated enlargement and consider
$H = (^*\C)^\eta$, $\eta\in{^*\N}$ (c.f. [Al86]). ${\cal A}$ is the
set of all internal partial functions $f\colon\dom f\to {^*\R}$,
$\dom f$ a hyperfinite $ON$--system in $H$ and $f\le g$,
iff as in section 2.1 $\tilde f\subseteq\tilde g$. Then by the proof
of lemma 3 internal observables correspond to symmetric and everywhere
defined linear mappings $L$ on $H$. If the norm of $L$ is finite,
then $L$ defines a bounded self adjoint mapping $\hat L$ on the
nonstandard hull $\hat H$ of $H$. If conversely $A$ is a bounded
self adjoint mapping on the standard Hilbert space $K$ and $H$ is
S--dense in $^*K$ (i.e. each $x\in ~^*K$ is infinitely close
to some $y\in H$), then for some bounded $L$ on $H$ (i.e. an observable of
${\cal A}$) $A$ is the restriction of $L$ to the standard elements. Thus
bounded quantum theory is obtained in principle
by means of the following modifications of its finitary variant:

\item{(i)} The reals are enlarged to a hyperreal number field;

\item{(ii)} finiteness is weakened to hyperfiniteness.
\msk

The philosophical implications of (ii) have been investigated by {\smc
Vop\^enka} [Vo79], in whose view hyperfinite infinity describes {\it ``the
phenomenon of infinity in accordance with our experience, i.e., as a
phenomenon involved in the observation of large, incomprehensible sets".}
To obtain quantum theory, that phenomenon is added to the finitary framework at
the level of the data $f\in{\cal A}$.
In order to represent such data within
an internal theory, the modification (i) is needed,
whereby the real numbers are
supposed to result from empirically realizable measurements and the
hyperreals and nonstandard states are theoretical entities. This folklore
interpretation is suggested by {\smc Abu'l
Hudhayl}'s ($9^{th}$ century AD) explanation of
{\smc Zeno'}s paradox of the arrow ({\smc Jammer} [Ja74], p. 259).
\msk

\Lemma {\it In $ZFA$, if $A: \dom A\to H$ is a finitary mapping and $K$ is a
invariant linear manifold in $\dom A$, then the restriction $B = A/K$ is
finitary.}
\msk

\Proof. The condition $ \dom A = \span EV(A)$ is equivalent to the assertion,
that for all $x \in \dom A$ also $Ax \in \dom A$ and $ \{ A^n x: \! n \in
\omega \} $ is finite dimensional. [ If $x = \sum _{i \in k} x_i$ and $Ax_i =
\lambda _i x_i$, then $A^nx = \sum _{i \in k} {\lambda _i}^n x_i$ and $\dim \{
A^n x: \! n \in \omega \} \leq k$. Conversely, if $S = \span \{ A^n x: \! n
\in \omega \} $ is finite dimensional, then by diagonalization $x = A^0 x\in S
= \span EV(A/S) \subseteq \span EV(A)$.] These conditions are inherited by
$A/K$.
\eop
\msk

If the invariant manifold $K$ is closed in $ \dom A$, then
we let $P$ be the orthogonal projection onto $K$. In $ZFA$, $Px$ is the
unique $y\in K$ such that $\Vert x-y\Vert = \inf\{\Vert x-z\Vert : z\in
K\}$. $Px$ exists, if $K$ is {\it Cantor
complete} (the intersection of a nested sequence of closed sets $C_n$ whose
diameters $\hbox{\rm diam} C_n$ converge to zero is nonempty.)
[In view of the proof in [Du57], p. 248--249,
$\hbox{\rm lim}_{\epsilon \to 0}\hbox{\rm diam} K_{\epsilon} = 0$,
$K_{\epsilon} = \{ y\in K:\ \Vert x-y\Vert\le\epsilon +
\inf\{\Vert x-z\Vert : z\in K\}\}$. Hence $\{ Px\} = \cap _{\epsilon > 0}
K_{\epsilon }$.] The existence of orthogonal projections on Hilbert spaces
depends on the axiom of choice, if no further completeness assumptions are
added (lemma 22.) Therefore some authors (c.f. [Mo91]) require
stronger completeness properties in the definition of a Hilbert space.
We denote by
$K^\bot = \{x\in H: x\bot K\}$ the orthogonal complement. Then
$A"(\dom A\cap K^\bot)\subseteq K^\bot$ [pick $x\in K, y\in
K^\bot\cap\dom A$; then $<x, Ay> = <Ax,y> = 0$, since $y\bot Ax\in K$]. If
$Ax = \lambda x $ then $Px\in EV(A)$ [ $A(Px) - \lambda (Px)
= A(Px - x) - \lambda (Px - x)\in K\cap K^\bot$.] Hence $K \cap EV(A) =
EV(A/K) = P"EV(A)$.
\msk

In $ZFA$, the restriction of a symmetric map $A$ to its domain is closed as a
mapping from $\dom A$ to $H$. Hence in $ZFA$ each eigenspace of a finitary map
is a closed and invariant subspace of $\dom A$ and moreover, quantum like maps
are closed.
\msk

\Lemma {\it In $ZFA$, the ideals $j(A_i)$ of the finitary mappings $A_1,\cdots
A_n$ are commeasurable, if and only if the mappings $A_i$ commute and their
domains are equal}.

\Proof. As follows from the proof of lemma 3 and 1.4., $j(A_i)$ are
commeasurable, iff $EV(A_i)\subseteq\span\bigcap\{EV(A_i): 1\le i\le n\}$. This
implies $\dom A_1 = \cdots = \dom A_n$ and $A_i\cdot A_j = A_j\cdot A_i$.
Conversely, let $A\cdot B = B\cdot A$ for $A, B$ finitary and let $\dom A =
\dom B$. If $\mu$ is an eigenvalue of $B$ and
$K_\mu = \hbox{ker} (B-\mu)$, then $AB
= BA$ implies $A''K_\mu\subseteq K_\mu$. We let
$C_\mu = A/K_\mu$. $C_\mu$ is finitary by lemma 4. Then $K_\mu = \span EV
(C_\mu)$ and since
$EV(C_\mu)\subseteq EV(A)\cap K_\mu\subseteq EV(A)\cap EV(B)$ and
$D = \dom A = \dom B$ is a
direct sum of spaces $K_\mu$ it follows, that $\span(EV(A)\cap EV(B)) = D$.
\eop
\msk

Commutativity alone does not suffice for commeasurability [$A = 0/ \{ 0 \} $
commutes with $B$, iff $B$ is one to one, for $BA = A$ and $AB = 0/ \ker
B$.]
\msk

If $F: {\R}^n\to {\R}$ is a function and $jA_i$ are commeasurable,
then $\hat F(jA_1,\cdots, jA_n)$ corresponds to the operator $A =
F(A_1,\cdots,A_n)$ such that $\dom A = \dom A_i$ and for $x\in\bigcap_i
EV(A_i)$, $A_ix = \lambda_ix$, $Ax = F(\lambda_1,\cdots\lambda_n)\cdot x$.
\msk

In $ZFA+$ "{\it Dedekind finite subsets of $\R$ are finite}"
a function $f:{\R}\to{\R}$ is admissible for the
class of all finitary observables on $\ell_2(\omega)$, if and only if it is
finite to one. [For if $f(\lambda_i) = \lambda$, $i\in\omega$ and $\lambda_i$ a
sequence of distinct reals, then the finitary operator $Ae_i = \lambda_ie_i$,
$e_i$ the canonical $ON$ base of $\ell_2(\omega)$, $\dom A_i =
\span\{e_i:i\in\omega\}$, corresponds to a finitary observable, but $\hat f(A)
= \lambda/ \dom A$ is strictly included in the finitary $\map\lambda \
(\lambda(x)= \lambda\cdot x, x\in\ell_2(\omega))$ and thus $\hat f(A)$
corresponds to an ideal which is not maximal]. In $ZFA$ it follows by the same
argument, that if $jA$
is a finitary observable, then $\hat f(A)$ corresponds to a finitary
observable for all everywhere defined Borel--functions $f$, iff $\dom A = H$.
Hence quantum like observables have the same functional calculus as the
observables of quantum theory.
In $ZFC$ this implies, that $A$ is algebraic, as follows
from the closed graph theorem and the following result due to {\smc
Kaplansky} [Ka54]: A bounded operator on a Hilbert space $H$ is algebraic, if
and only if its finite dimensional invariant subspaces cover $H$.
\msk

\Lemma  {\it In $ZFC$, a self adjoint operator is finitary, iff it is
bounded and algebraic $(p(A) = 0$ for some nonzero polynomial $p$).}
\msk

\Proof. If $A$ is finitary, then $\dom A =
\oplus_{\lambda\in\sigma_p(A)} \hbox{ker}(A-\lambda)$
(= linear direct sum of the
orthogonal system of eigenspaces) and
$\hbox{im} A\subseteq\dom A$. Since $A$ is
self--adjoint, $\dom A$ with the norm $\Vert x\Vert^2_1
= \Vert x\Vert^2 + \Vert Ax\Vert^2$ is a Hilbert space (c.f. [Du63], p.1225)
and by the closed graph theorem, $A$ is bounded. {\smc Kaplansky's} [Ka54]
theorem, applied to $A$
on this space, implies that $A$ is algebraic on $\dom A$ and since $\dom A$ is
dense for a self adjoint operator $A$, $A$ extends to an unique bounded [since
the spectrum $\sigma(A)$ is finite,
$A$ is bounded in the norm of $H$] and algebraic
operator on $H$. If conversely $A$ is bounded and algebraic, then in view of
{\smc Kaplansky's} theorem $H = \span (EV(A))$, whence $A$ is
finitary.
\eop
\msk

In $ZFC$, quantum like observables are empirically realizable in a weak sense.
For as has been shown by {\smc Zeilinger} and his collaborators [Re94], the
time evolution of algebraic Hamiltonians may be simulated by photon--splitting
experiments. [Technically, this result is an analogy of {\smc Jacobi}'s $QR$--
decomposition of a symmetric matrix into a diagonal matrix and two dimensional
rotations.]
\msk\msk

\no {2.3. Russell's socks}.
In $ZFC$, there is an abundance of non--effective pure states in an
infinitely dimensional Hilbert space. Since {\smc Kaplansky}'s theorem depends
on $AC$ ({\smc Brunner} [Br86]), these states are responsible for the
restrictions on the quantum like observables. We therefore
investigate finitary observables in models of $ZFA$ minus the axiom of choice.
The most interesting
spaces for our purpose are the {\it locally
sequentially compact} ones (the unit sphere is sequentially compact).
\msk

\Lemma  {\it In $ZFA$, a Hilbert space $H$ is locally sequentially compact, iff
each bounded symmetric operator $A\colon H\to H$ is finitary.}
\msk

\Proof. If $H$ is locally sequentially compact, then each $x\in H$ is
contained in a finite dimensional invariant subspace $F$ of $A$ ({\smc
Brunner} [Br86], 4.3). By diagonalization, $F = \span(EV(A\mid F))$, whence
$A$ is finitary.  If $H$ is not locally sequentially
compact, then it contains a copy of $\ell_2(\omega)$ {\smc (Brunner} [Br86],
2.1). Since $\ell_2(\omega)$ is second category, {\smc Kaplansky's} theorem
applies, whence no symmetric bounded operator $A$ on $\ell_2(\omega)$ with
infinite spectrum is finitary. Its extension $A\circ P$ to $H$ ($P$ the
projection onto the Cantor complete subspace $\ell_2(\omega)$)
is a symmetric bounded operator on $H$
which does not correspond to a finitary observable (c.f. lemma 4).
\eop
\msk

\Lemma  {\it In ZFA, if A is a quantum like map on the Cantor complete
Hilbert space H such
that each eigenspace of A is finite dimensional, then H is locally
sequentially compact.}
\msk

\Proof. For $\lambda \in \R$ we let $P_{\lambda }$ be the orthogonal
projection onto $\ker (A - \lambda)$ and define, for $x\in H$, its support as
$s(x) = \{ \lambda \in \R :\! P_{\lambda }x \not= 0 \} $. Then for each
sequence $\< x_n:\! n\in \omega \>$ in $H$ the set $S = \cup \{ s(x_n):\! n\in
\omega \}$ is finite. For otherwise there exists an infinite partition $S_k
\not= \emptyset$, $k\in \omega $, of $S$ (an observation due
to {\smc D. Pincus}; c.f. [Br82], corollary 2.2.5.) We set $K_k = ${\bf
cl}$\oplus _{\lambda \in S_k}\ker (A - \lambda )$ and let $Q_k$ be the
orthogonal projection onto $K_k$. Then for each $k\in \omega $ there is a
least index $n(k)$ such that $y_k = Q_kx_{n(k)} \not= 0$, since $Q_k \ge
P_{\lambda }$ for $\lambda \in S_k$. We set $y = \sum _{k \in \omega
}{1\over {k+1}} \cdot {{y_k}\over {\Vert y_k\Vert }} $.
Then $y \in H$, but $y\notin
\span EV(A)$. For if $y \in \oplus _{i \in n}\ker (A - \lambda _i)$ and
$S_k\cap \{ \lambda _i:\! i\in n\}  = \emptyset $, then $Q_k" \oplus_{i\in
n}\ker (A - \lambda _i) = \{ 0\} $, but $Q_ky = {1\over {k+1}} \cdot
{{y_k}\over {\Vert y_k\Vert }} \not= 0$ [$Q_ky_h = 0$ for $k\not= h$, since
distinct eigenspaces of $A$ are orthogonal.] As $y\notin \span EV(A)$ is
impossible, if $A$ is quantum like, $\span \{ x_n:\! n\in \omega \} \subseteq
\oplus _{\lambda \in S}\ker (A - \lambda ) $ is finite dimensional, whence $H$
is locally sequentially compact ([Br86], lemma 2.1.)
\eop
\msk

There are also set theoretical restrictions on Cantor complete Hilbert
spaces $H$ which admit quantum
like mappings $A$ whose eigenvalues have
finite multiplicities, only. In this case
for each infinite $D\subseteq H$ the powerset ${\cal P}(D)$ is Dedekind
infinite. [For otherwise by the above mentioned result due to {\smc Pincus}
$\{ s(x):\ x\in D\} \subseteq [\R]^{< \omega}$ is finite. Therefore for some
$E\in [\R]^{<\omega}$ $B = \{ x\in D:\ s(x) = E\}$ is
infinite. But $B$ induces an infinite subset of $\C ^m$, $m = \sum_{\lambda
\in E}\dim (\ker (A-\lambda ))$, with a Dedekind finite
powerset, contradicting the result of {\smc Pincus}.]

\msk
The existence of such spaces is motivated by thought experiments about
{\smc Russell}'s socks, the standard example for a failure of $AC$. In our
sense they
form a sequence of pairwise disjoint two element sets $P_n = \{a_n, b_n\},
n\in\omega$, which is a counterexample to the { \it principle of
partial dependent choices} [$PDC:$ there exists
a sequence $F_k: C_{n_k}\to P_{n_k}$ of
functions, where $C_{n_k}$ is the set of choice functions on $\< P_i : i\in
n_k\> , n_k<n_{k+1})$ for $k\in\omega$].
Thus it is not possible to single out a
sock of the $n-th$ pair in an uniform way, even if one could distinguish the
previous socks. We view $\{a_n, b_n\}$ as an assembly of identical
noninteracting spin $1\over 2$ particles which obey the {\smc Fermi--Dirac}
statistics (c.f. {\smc Jauch} [Ja68], pp. 249 -- 287). Its Hilbert
space $H_n = \span \{e_1(a_n)\otimes e_2(b_n) - e_2(a_n)\otimes e_1(b_n)\}$ is
isomorphic with $L_n = \{x\in\ell_2\{a_n, b_n\}: x(a_n) + x(b_n) = 0\}$. The
family of all socks is viewed as the compound system of the distinguishable
assemblies. Then the failure of $PDC$ might be considered a set theoretical
implementation of the physical independence of these particles.
Their {\smc Fock} space is $F = \oplus_{N\in\omega}\quad\otimes_{n\in N}H_n$.
Here we may replace $H_n$ by
$L_n$, since the isomorphism does not depend on the ordering of $P_n$. The
finite tensor product $\otimes_{n\in N}\quad L_n$ is $T_N$, where
$$T_N = \{X\in\ell_2(C_N) : X(\phi) = (-1)^mX(\psi),\quad if\quad \vert\{i\in
N:\phi(i)\not=\psi(i)\}\vert = m\}$$
and $C_N$ is the set of choice functions
$\phi: N\to\cup_{n\in N}P_n, \phi(n)\in P_n\hbox { for } n\in N$. This follows
from the definition $\tau_N : \Pi_{n\in N}L_n\to T_N, \tau_N({\bf x}) = X$,
where ${\bf x}= <x_n: n\in N>$ and $X(\phi) = \Pi_{n\in
N}x_n(\phi(n))$ is the product in ${\C}$. The function $\tau_N$ is a
multilinear onto map such that
$$\eqalign{%
<\tau_N({\bf x}), \tau_N({\bf y})>_{T_N} &= \sum_{\phi\in C_N}\prod_{n\in
N}x_n(\phi(n))y_n(\phi(n))^-\cr
&= 2^N\cdot \prod_{n\in N}x_n(a_n)y_n(a_n)^-\cr
&= \prod_{n\in N}<x_n, y_n>_{L_n}.\cr}$$
While $T_N$ is isomorphic to ${\C}$, the direct sum of the finite tensor
products is not $\ell_2(\omega)$, since the isomorphism depends on the
ordering. Rather the {\smc Fock} space is isomorphic with $T = \{X\in\ell_2(C)
: X(\phi) = (-1)^mX(\psi)$, whenever $\dom\phi = \dom\psi = N
$ and $\vert\{i\in
N:\phi(i)\not=\psi(i)\}\vert = m$, $N$ and $m$ in $\omega\}$, where $C =
\cup\{C_N\colon N\in\omega\}$.
\msk

$MC^\omega$ is the {\it countable multiple choice axiom} (for each sequence of
non\-emp\-ty sets $S_n, n\in\omega$, one may
choose a sequence of nonempty finite
subsets $E_n\subseteq S_n, n\in\omega)$. In $ZF$, $MC^\omega$ depends on $AC$,
but it seems not exclude {\smc Russell}'s socks [{\smc Sageev} [Sa75]
constructs a model of $MC^{\omega}$ where $AC^{\omega}$ fails. His proof
appears to extend to $not\ PDC$. In $ZFA$ the corresponding
result is a triviality by section 2.4.]
$MC^{\omega}$ implies, that Cantor completeness is equivalent with
completeness ([{\smc Br}83], lemma 3.4.)
\msk

\Lemma {\it In $ZFA+$"there are } {\smc Russell}{\it 's socks",
$T$ (and thus $F$) is a locally sequentially compact
Hilbert space (norm of $\ell_2$). There are nonalgebraic bounded
quantum like
operators on $T$. In $ZFA + MC^{\omega }+ $"there are }{\smc Russell}{\it 's
socks" each system of linearly
independent vectors in $T$ is finite.}
\msk

\Proof. The first and third parts follow as in {\smc Brunner} [Br90], proof of
example 4.4. For $X \in T$ we set $s(X) = \{ \phi \in C: \! X(\phi) \ne 0\} $.
We let $ \prec $ be a lexicographic ordering on $\C$ such that $ a \prec 0$
implies $0 \prec -a$. If $X_n$, $ n \in \omega$, is a sequence in $T$, then
$S = \cup \{ s(X_n): \! n \in \omega \} $ is finite. For otherwise we define a
$PDC$ function $F$ as follows. $N = \{ n \ge 1: \! S \cap C_n \ne \emptyset
\}$ would be infinite. If $n \in N $, let $m$ be the least index such
that $s(X_m) \cap C_n \ne \emptyset $. Then $X_m(\phi ) \ne 0$ for all
$\phi \in C_n$. For $\psi \in C_{n-1}$ we set $F_n(\psi ) =  \phi (P_n)
\in \{ a_n,b_n \} $, if $ \psi \subseteq \phi \in C_n $ and $X_m(\phi )
\succ 0$.
\msk

It follows, that $X_n$ is a sequence in the finite dimensional space
$\ell_2 (S)$. This proves completeness and locally sequentially
compactness.
\msk

Since each real diagonal operator $D$ on $\ell_2(\omega)$ induces
a symmetric bounded operator $A$ on $T$ with the same spectrum, $(AX)(\phi) =
d_nX(\phi)$, if $\phi\in C_n$ and $De_n = d_ne_n$ for the unit vector $e_n$ of
$\ell_2(\omega)$, the existence of counterexamples to {\smc Kaplansky'}s
theorem follows from lemma 7.
\msk

If $D \subseteq T$ is a system of linearly independent vectors, then in
view of the locally sequential compactness of $T$ $[D]^{< \omega }$ is
Dedekind finite ([Br86], lemma 2.1).
$MC^{\omega }$ implies, that $D$ is finite. [As in lemma 10 below, $S =
[[D]^{<\omega}]^{<\omega}$ is Dedekind finite with an infinite partition $X_n
= [D]^n$ of $[D]^{<\omega}$. If $\emptyset\neq E_n\subseteq X_n$
is finite, then since $E_n\cap
E_m = \emptyset$ for $n\neq m$ $S$ contains the infinite sequence $\<E_n:\
n\in\omega\>$. This is improved in lemma 22.]
\eop

$F$ and also the simpler space $L = \oplus_n L_n$ are counterexamples to
several assertions of Hilbert space theory in $ZFC$.
\msk

\item{(i)} Both spaces admit no infinite $ON$ system. Thus it is not possible
to choose a mode of observation (in the sense of {\smc Bohr}'s complementarity
interpretation) by choosing an $ON$ base. Moreover, there is no Hamel base,
either. Therefore the {\it multiple choice axiom} $MC$ in $ZFA$ does not imply
the existence of bases, although it is known to be a consequence thereof
([Ru85], p. 119.) Similar vector spaces have been constructed by {\smc
L\"auchli} [La62]. His construction may be modified so as to yield locally
sequentially compact Hilbert spaces of the second category such that all
closed and everywhere defined linear operators are algebraic (improved in
lemma 21).

\item{(ii)} The {\smc Riesz} representation theorem for continuous
linear functionals is invalid, since
the duals of $F$ and $L$ differ from $F$ and $L$, whence the
notion of a selfadjoint operator does not make sense and the usual approach to
quantum theory fails. As the {\smc Hahn--Banach} theorem is a consequence of
$MC$ ({\smc Pincus} [Pi72]), in $ZFA$ {\smc Riesz}' theorem
does not follow from the {\smc Hahn--Banach} theorem. We note, that locally
$MC^{\omega} + \exists \hbox{{\smc Russel}{\it 's socks}}$ suffices for such a
conclusion. [For then $F$ is Cantor complete and each linear functional
attains its norm on the unit sphere. Now an argument due to {\smc Ishihara}
[Is89] proves $HB$ for $F$; c.f. [Mo91].]
\msk

If {\smc Russell}'s socks satisfied the {\smc Einstein--Bose} statistics,
then their {\smc Fock} space $\ell_2(\omega)$
would not admit nontrivial quantum like
observables. We do not, however, attempt to
resolve the physical nature of {\smc Russell}'s socks.
\bsk

More straightforward examples of nontrivial quantum like
observables are obtained
from {\smc Dedekind} sets $D$ (infinite, Dedekind finite sets of reals).
For $(Ax)(d) = d\cdot x(d)$ is a quantum like mapping on $\ell_2(D)$ with one
dimensional eigenspaces and point spectrum $\sigma_p(A) = D$. By contrast, $L$
admits quantum like mappings with countably infinite point spectra and finite
multiplicities.
\msk

\Lemma {\it In $ZFA$, if there is an infinite set $D$ such that its set of
finite subsets $[D]^{<\omega}$ is Dedekind finite, then there exist
self--adjoint quantum like nonalgebraic bounded observables on
$H = \ell_2([D]^{<\omega})$.}
\msk

\Proof. $H$ is locally sequentially compact
({\smc Brunner} [Br86]), since $[[D]^{<\omega}]^{<\omega}$ is Dedekind finite
[if $E_n\in[[D]^{<\omega}]^{<\omega}$ is a sequence, we set
$\cup E_n\in[D]^{<\omega}$ and in view Dedekind finiteness $\cup\{\cup E_n :
n\in\omega\} = E\in[D]^{<\omega}$, whence $E_n\in{\cal PP}(E)$ is a finite
sequence]. However, ${\cal P}([D]^{<\omega})$ is not Dedekind finite $[X_n =
[D]^n, n\in\omega]$, whence on $H$ there exist real diagonal operators with
infinite point spectra. Lemma 7 concludes the proof.
\eop
\msk\msk

\Lemma In $ZFA+MC^{\omega}$ any quantum like mapping $A$ on
$H = \ell_2(D)$ with finite dimensional eigenspaces is algebraic.
\msk

\Proof. By lemma 8 and [{\smc Br}83], lemma 3.4, $H$ is locally
sequentially compact, whence by [Br86], lemma 2.1, $[D]^{<\omega}$ is Dedekind
finite and as in lemma 9 $D$ is finite.
\eop
\msk\msk

\no {2.4. Weglorz models}. As a first step in relating quantum like
observables in set theories without $AC$ to quantum theoretical observables in
a $ZFC$ universe $V$, we let the finitary observables live in a world which is
interpreted within $V$ as a permutation model. In the sequel we shall
review the construction of permutation models of
$ZFA$.
\msk

Inside the real world $V$ which satisfies $ZFC$ we construct for some $X\in V$
a $ZFA + AC$ model $V(X)$, following an idea by {\smc J.~Truss}. In
$V(X)$ the set of atoms will correspond to $X$. The definition is by
recursion: $V(X) = \cup\{V_\alpha\colon\alpha\in On\}$, where $V_o =
X\times\{0\}$
and $V_\alpha =\{(A,\alpha)\colon\alpha\hbox{ minimal}$, such that
$A\subseteq\cup\{V_\beta\colon\beta\in\alpha\}\}$. If $x$ and $y$ are objects
in $V(X)$, then $x\simeq y$ in $V(X)$, iff $x = y$ in the real world, and
$x\tilde\in y$ in $V(X)$, iff $y = (A,\alpha)$ for some $\alpha>0$ and
$x\in A$.
\msk

The general construction of a {\smc Fraenkel--Mostowski} model is as follows. A
group $(G,\cdot)$ is given.
If $d\colon G\to S(X)$ is an injective homomorphism into the
symmetric group over $X$, then $d$ is recursively extended to $\hat d$ on all
of $V(X)\colon \hat d g(x,0) = (dg(x), 0)$ and $\hat dg(A,\alpha) = (\hat
dg''A,\alpha)$ for $\alpha\ge 0$. If $\underline G$ is a $T_2$ group topology
on $G$
which is generated by a n.h. base at $1$ which consists of open subgroups,
then an object $x\in V(X)$ is fixed relative to $\underline G$, if the
stabilizer of $x$, $\hbox { stab }x = \{g\in G\colon \hat dg(x)\simeq x\}$ is
open. $PM$ consists of the hereditarily fixed objects; $PM = \{x\in
V(X):$ all elements in the $\tilde \in$-- transitive closure of $\{x\}$ are
fixed relatively to $\underline G\}$.
\msk

$(PM,\tilde\in)$ is a model of $ZFA$. If $AC$ holds in the real world, then
$PM$ satisfies $AC$, iff $\underline G$ is discrete.
\msk

For example, the second {\smc Fraenkel} model $FM$ is
generated by the group $G = \Z^\omega_2$ with the product topology and
the natural action on $X = \cup\{\Z_2\times\{n\}\colon n\in\omega\}$. We
recall from {\smc Jech} [Je73], that $FM$ admits {\smc Russell}'s socks and
thus nontrivial quantum like observables by lemma 9 and $FM$ satisfies the
{\it multiple choice axiom} $MC$ ([Ru85], p. 74.)
\msk

The following speculations serve as a motivation for the next section, where
we shall investigate effective descriptions of the quantum theoretical
observables.
If an element $x$ of the $ZFC$ world $V$ has an isomorphic Dedekind finite copy
$y$ in a permutation model $PM$, then $y$ might be
considered as an {\it effective
representation} of $x$. We illustrate this by the following
example.
\msk

In $V$ we let $\Omega$ be an atomic Boolean algebra with $X$ the set of its
atoms. Then $\Omega$ is isomorphic with a set algebra $X_\Omega\subseteq
{\cal P}(A)$
of $V(X)$, $A$ the set of atoms in the sense of sets. We let $\Aut(\Omega)$ be
the group of all Boolean algebra automorphisms of $\Omega$. Since $\Omega$ is
atomic, the automorphisms are induced by bijections on the set of atoms,
whence $\Aut(\Omega) = \Aut X_\Omega<S(A)$, the symmetric group. Since
$X_\Omega\supseteq[A]^{<\omega}$, the topology which is generated by the n.h.
base at $id$ of the
subgroups $\hbox{ stab }(x), x\in X_\Omega$, is $T_2$ and it
defines a {\smc Fraenkel}--{\smc Mostowski} model $PM_\Omega$. In view of its
similarity to a construction by {\smc Weglorz} [Wg69] we baptize it {\smc
Weglorz} model.
\msk

\Lemma {\it In $PM_\Omega$, $X_\Omega = {\cal P}(A)$.}
\msk

\Proof. This result is due to {\smc Weglorz} [Wg69]. His proof works with
$SF(A)$, the group of all finite permutations on $A$. But since
$SF(A)<\Aut(\Omega)$, it is valid for $PM_\Omega$, too.
\eop

\Lemma {\it In $PM_\Omega$ ${\cal P}(A)$ is Dedekind finite.}
\msk

\Proof. We may assume, that $\Omega$ is infinite. We let $\<E_n: n\in\omega\>$
be a sequence of subsets of $A$ in $PM_\Omega$. Then for some sets
$x_1,\cdots,x_k\in X_{\Omega}$
and all $n\hbox { stab}E_n\supseteq\bigcap_{1\le i\le k} \hbox{ stab }
x_i$. If $B$ is the finite Boolean algebra which is generated by the $x_i$ and
$A$, then we may assume $\{x_1\cdots,x_k\} = B$. The atoms of $B$ form a
partition of $A$. We
show, that $E_n\in B$, whence the sequence $\<E_n: n\in\omega\>$ is
finite. Pick an atom $y\in B$ and $a\in E_n\cap y$. For $b\in y$
we let $\pi\in SF(A)$ be the transposition $\pi = (a; b)$. Then $\pi$ leaves
invariant all the atoms of $B$ (partition) and therefore the other elements of
$B$, too, whence $\pi\in\hbox { stab}E_n$ and $b\in E_n$. Hence $y\subseteq
E_n$ and $E_n$ is a finite union of the atoms $y\in B$ such that $y\cap
E_n\not=\emptyset$, whence $E_n\in B$.
\eop

As follows from lemma 13, the theory of the atomic Boolean algebras of $V$ has
a nontrivial effective content. By contrast, an infinite rigid Boolean
algebra does not admit a Dedekind finite representation $B$ in any
permutation model $PM$ [In the notation of [Br95] $PM(B)$ satisfies $AC$ by
rigidity, whence $B$ would be finite by Dedekind finiteness.]
Moreover
in all {\smc Weglorz} models there are nontrivial quantum like
observables, if $\Omega$ is infinite (by lemmas 10 and 13).
\msk

Since in $PM_\Omega$ the set $[A]^{<\omega}$ is Dedekind finite, $PM_\Omega$
does not
satisfy $MC^\omega$, unless $X_\Omega$ is finite, whence $FM$ is distinct from
these models. Moreover, if $X_\Omega$ is infinite, then in $PM_\Omega$ there
is no choice function on $[A]^2$ [for otherwise, if $f$ is a choice function
and $\hbox{ stab } f
\supseteq\bigcap\{\hbox{ stab }(b): b\in B\}$, $B$ a finite Boolean algebra,
then for an infinite atom $y\in B$, $\{a,b\}\in[y]^2$ and
$\pi = (a;b)\quad f\{a,b\} =
f\pi\{a,b\} = \pi f\{a,b\}$, contradicting the definition of $\pi]$. Hence the
ordered {\smc Mostowski} model ({\smc Jech} [Je73], p. 49) is different from
$PM_\Omega$. In general, $PM_{\Omega}$ is different from the original model of
{\smc Weglorz}. In {\smc Weglorz}' construction the Boolean algebra
${\cal B}_{\Omega} = {\cal P}(A)/ [A]^{<\omega }$ is wellorderable. In
$PM_{\Omega}$ it is infinite, but Dedekind finite, if in $V$ $X_{\Omega} =
{\cal P}(A)$ is infinite. [If $B\subseteq{\cal P}(A)$
is a finite Boolean algebra $(A\in B)$ and ${\cal E}\subseteq{\cal B}_\Omega$
is a finite set such that
$\hbox{ stab } {\cal E}\supseteq\bigcap\{\hbox{ stab } x : x\in B\}$,
then for each $E\in{\cal E}$ there is a $P\in B$ such that the symmetric
difference $E\triangle P$ is finite. For otherwise for some infinite atom (the
atoms of $B$ form a finite partition of $A$)
$x\in B$ and some $E\in{\cal E}$ the sets $E\cap x$ and
$x\setminus E$
are infinite, whence we may define bijections $\pi_n$ such that $\pi_n"x = x$,
$\pi_n(y) = y$ for $y\in A\setminus x$ and $\pi_n(x\cap E)\triangle\pi_m(x\cap
E)$ is
infinite for $n\not= m$ in $\omega$. Then
$\pi_n\in \hbox{ stab } {\cal E}$, but the set of equivalence classes
$\{[\pi_n E] : n\in\omega\}\subseteq{\cal E}$ is infinite, a contradiction.
Hence each sequence of finite subsets of ${\cal B}_\Omega$ which is supported
by $B$ is contained in the finite set $\{[x] : x\in B\}$.]
\msk\msk

\no {2.5. Benioff extension}. The previous lemmata provide us with several
examples of quantum like observables in permutation models $PM$. In
order to compare their theory with quantum mechanics, we apply an idea due to
{\smc Benioff} [Be76] and investigate the model $PM$ from outside, where $AC$
holds. In $PM\subseteq V(X)$ we let $A$ be a quantum like mapping on
the Hilbert space $H$. In $V(X)$, $H$ is an inner product space, whence its
completion $\tilde H$ is a Hilbert space. For example, $L\in FM$ (section 2.3)
in $V$ is isometrically isomorphic with the space of polynomials
$K = \C [x]\cap {\cal L}_2]0,1[$ and
$\tilde L$ with ${\cal L}_2]0,1[$. The quantum like mappings $A$ on $L$
correspond to a proper subset of the family of all linear symmetric mappings
on $K$. [The multiplication $(Qf)(x) = x\cdot f(x)$ on $K$ does not correspond
to any mapping on $L$. For if in $V$ $Q$ is unitarily equivalent to a linear
mapping $A\in PM$ on $H\in PM$, i.e. $A\cdot U = U\cdot Q$ for some bijective
isometry $U:\  H\to K$ in $V$, then in $PM$ for $h\neq 0$ in $H$ the set
$\{ p(A)(h):\  p\in \C [x]\}$ is dense in $H$, since by [Ra73], p. 95 -- 96,
$Q$ does not admit a
nontrivial closed invariant subspace. Thus in $PM$ the inner product space $H$
is a separable Hausdorff space of the first category.
It is therefore wellorderable as a set by [Br83], lemma 2.2, whence
in $PM$ $H$ cannot be a Hilbert space.] We show, that there is a unique
selfadjoint extension $\tilde A$ of a quantum like
$A$ to $\tilde H$; if $A$ is unbounded, then $\dom\tilde A\not= \tilde H$.
\msk

\Lemma {\it In $ZFC$, if the linear mapping $A$ corresponds to a finitary
observable,
then its domain $\dom A$ is dense and there exists a unique selfadjoint
extension $\tilde A$ of $A$. If conversely $A$ is a selfadjoint operator such
that $\hbox{\bf cl}\span EV(A) = H$, then $A^\circ = A/\span EV(A)$
corresponds to a finitary observable.}
\msk

\Proof. Existence and uniqueness of $\tilde A$ follows from a calculation of
the deficiency indices $\dim\{x\in\dom A^*:A^*x = \pm ix\} = 0$. The
adjoint $A^*$ is
defined from a complete $ON$ system $B$ of eigenvectors of $A$ as the diagonal
operator $A^*b = Ab = \lambda_b\cdot b$ for $b\in B$ and $\dom A^* = \{x\in
H:\sum_{b\in B}\lambda^2_b\cdot\vert\<x,b\>\vert^2<\infty\}$;\  $A^* =
\tilde A$ (c.f. {\smc von Neumann} [vN29]).
\eop
\msk

As follows from standard results in spectral theory, in $ZFC$ $A^\circ$ is
defined for all bounded symmetric operators with a countable spectrum.
Another example are the density operators of quantum theory.
\msk

\Lemma {\it If in the permutation model $PM\subseteq V(X)$ the mapping $A:
H\to H$ on the Hilbert space $H$ is quantum like, then in $V(X)$
there exists a unique self adjoint extension $\tilde A$ of $A$ to the
completion $\tilde H$ of $H$; $\tilde A^\circ\supseteq A$. If $f: \R^n\to\R$
is an everywhere defined Borel function and in $PM$ the mappings $A_i$
correspond to commeasurable quantum like observables on $H$, then in
$V(X)$ the self adjoint mappings $\tilde A_i$ commute and $f(A_1,\dots
A_n)^\sim = f(\tilde A_1,\dots \tilde A_n)$}.
\msk

\Proof. It follows from {\smc Benioff} [Be76], that each $x\in EV(A)$ in $PM$
is an eigenvector of $A$, when viewed in $V(X)$ as an operator on the
prae--Hilbert space $H$.
Hence also in $V(X)$ $H = \span EV(A)$. In $V(X)$ it follows
from $AC$, that there is a complete $ON$--system $B\subseteq EV(A)\subseteq H$
of $\tilde H$. The proof of the preceding lemma shows, that in $V(X)$ $A^*$ is
the unique self adjoint extension $\tilde A$ of $A$. It is obvious, that
$\tilde A^\circ\supseteq A$. Commeasurability of quantum like
observables is
equivalent with commutativity by lemma 5, and the $A^*_i$ commute, if the
finitary $A_i$ do. Since then $f(\tilde A_1,\dots\tilde A_n)$ is a self adjoint
extension of the quantum like observable $f(A_1,\dots A_n),
f(A_1,\dots A_n)^\sim = f(\tilde A_1,\dots\tilde A_n)$ follows from
uniqueness.              \eop
\msk

$\tilde A$ is the {\it Benioff extension} of $A$. A self adjoint operator
which in the real world $V$ of $ZFC$
is unitarily equivalent to an operator of the form $\tilde A$ for some
quantum like mapping $A$ in some permutation model is called
{\it intrinsically effective}. ($A$ on $H$ and $B$ on $K$ are unitarily
equivalent, if for some bijective isometry $U\colon H\to K, UA =BU$).
If $T$ is a self adjoint operator on the Hilbert space $K$, then $T$ is
intrinsically effective, iff $\dom T^\circ $
is dense in $K$ (c.f. the proof of
lemma 16 below), iff $T^\circ $
corresponds to a finitary observable on $K$. The
correspondence between $T$ and $T^\circ $
is delusive, however, since in general in view of the discussion preceding
lemma 6 $f(T)^\circ \neq f(T^\circ )$,
whence this correspondence does not
respect the functional calculi. On the other hand, as a consequence of lemma
20 below  the Benioff extension respects the functional calculus of quantum
theory in a strong sense.
\msk

\Lemma {\it In the real world $V$ of $ZFC$, if $T$ is an intrinsically
effective self adjoint operator on the Hilbert space $K$ and $t\in \dom
T^\circ$, then there is a
permutation model $PM$ of $ZFA + MC$ which contains a quantum like
mapping $A$ on a locally sequentially compact Hilbert space $L$, such that
for some bijective isometry $\tilde U:\tilde L\to K$in $V$ $\tilde U^{-
1} t\in L$ and in $V$ $T$ is unitarily equivalent with $\tilde A$,
$T\tilde U = \tilde U\tilde A$.}
\msk

\Proof. Since $T$ is intrinsically effective, $T^\circ$ contains some finitary
mapping on a dense submanifold of $K$, whence $\span EV(T)$ is dense in $K$. In
$V$ we choose an $ON$--base of $K$ which consists of eigenvectors $k_\alpha,
\alpha\in\kappa$, of $T$; $Tk_\alpha = \lambda_\alpha\cdot k_\alpha$.
If for some eigenvectors $k_i$, $i\in n$ of $T$ $t = \sum_{i\in n} t_ik_i$,
then we let the $ON$--base contain these $k_i$ and let
$H$ be its linear span. In the model $PM$ which is generated by the
compact topological group $G = \Z_2^\kappa$, and thus satisfies $MC$, we
repeat the construction of the Hilbert space $L$ which is now related to a
transfinite sequence of {\smc Russell}'s socks $P_\alpha = \{a_\alpha,
b_\alpha\}, \alpha\in\kappa$; $L =
\{x\in\ell_2(P)\colon\forall\alpha\in\kappa\colon x(a_\alpha) + x(b_\alpha) =
0\}$ where $P = \cup_{\alpha\in\kappa}P_\alpha$. On $L$ we define a linear
mapping $A\in PM$ with $\dom A = L$ through the clause $(Ax)(p) =
\lambda_\alpha\cdot x(p)$, if $p\in P_\alpha$. Then in $PM$\  $L$ is locally
sequentially compact (c.f. lemma 9) and in $V$ there is a bijective isometry
$U$ between $L$ and $H$ which extends to an isometry $\tilde U$ between
$\tilde L$ and $\tilde H = K$; by the definition of $L$ $U^{-1}(t)\in L$.
$PM$ does not depend on $t$, but $U$ does.
Since $T/H = UAU^{-1}, T = \tilde U\tilde A\tilde U^{-1}$
and $\tilde A,\ T$ are unitarily equivalent.
\eop
\msk

Locally sequentially compactness is inevitable, for by
lemma 8, if $T$ is the Benioff
extension of $A$ in $PM$ which satisfies $MC$
and the multiplicity of all eigenvalues of $T$ is
finite, then in $PM$ $\dom A$ is locally sequentially compact [modulo an abuse
of notation, $\ker (A - \lambda ) = \dom (A)\cap \ker (T - \lambda)$ is finite
dimensional.] Lemma 16 indicates, that the notion of effectiveness which
corresponds to intrinsic effectivity is $ZFA+MC$ (instead of just $ZFA$). If
in the above proof the powers of $\Z_2$ are replaced by p--adic groups, then
finite axioms of choice are added (c.f. [Br90].) As follows from the proof of
lemma 16, a self adjoint operator $T$ on the Hilbert space $H$ is
intrinsically effective, iff its eigenvectors span a dense submanifold of $H$.
\msk

\Lemma {\it In $ZFC$, a bounded symmetric operator $T$ on a Hilbert space $H$
is intrinsically effective, iff it is unitarily equivalent to a multiplication
operator on a functional Hilbert space}.
\msk

\Proof. This result follows from {\smc Halmos}' characterization of the latter
property (c.f. [Ha82], problem 85), which holds, iff $H = {\bf {cl}} \span
EV(T)$. In view of the above proof this is equivalent with intrinsic
effectivity.
\eop
\msk

The notion of intrinsic effectivity may be generalized somewhat. We let $C(H)$
be a definition of a class of symmetric and {\it bounded} mappings on a
prae--Hilbert space $H$. We assume

\item{(i)} {\it absoluteness} in the following sense: If in the permutation
model
$PM\subseteq V(X)\  A\in C(H)$, $H$ a Hilbert space both in $PM$ and in
$V(X)$, then also in $V(X)\  A\in C(H)$ (c.f. lemma 15);
\item{(ii)} {\it finiteness}: In $ZFC$, if $H$ is a Hilbert space, then $C(H) =
C_A(H)$, the algebraic operators (c.f. lemma 6);
\item{(iii)} {\it invariance}: In $ZFA$, if $H$ is a Hilbert space, $A\in C(H)$
and the closed
subspace $K\subseteq H$ is an invariant subspace for $A$, then $A/K\in C(K)$
(c.f. lemma 4).

Since each $A\in C(H)$ is bounded, there is a unique extension $\tilde A$ of
$A$ to the completion $\tilde H$ of $H$. In $ZFC$ we thus may define the class
$C^{PM}(H)$ for a Hilbert space $H$ and a permutation model $PM$ as follows:
$A\in C^{PM}(H)$, iff for some Hilbert space $K$ in the sense of $PM$ up to
unitary equivalence $\tilde K = H$, $A/K\in PM$ and in $PM\  A/K\in C(K)$.
\msk

\Lemma  {\it In $ZFC$, each operator in $C^{PM}(H)$ is intrinsically
effective.}
\msk

\Proof. We assume $H = \tilde K$, $PM\models A\in C(K)$, pick  $k\in K$, let
$G$ be the stabilizer $\hbox{ stab }(k,K,A,C(K))$ and set $K_G = \{x\in K\colon
\hbox{ stab }(x)\supseteq G\}$. Since $G$ fixes the
topology of $K$, $K_G$ is closed in
$PM$. Since in $PM$ \  $K_G$ is wellorderable as a set, $PM$ contains all
$V(X)$--sequences of elements in $K_G$, whence $K_G$ is closed also in $V(X)$
as a subspace of $H$. Since $G$ fixes $A$ and $k\in K_G$, $\{A^nk\colon
n\in\omega\}\subseteq K_G$ (here $A^\circ = id$) and $orb(k) = {\bf {cl}}_H
\span \{A^nk\colon n\in\omega\}$ is closed both in $H$ and in $K$ and so
$orb(k)\subseteq K_G$. Moreover, by continuity, $orb(k)$ is invariant both for
$A$ and for $\tilde A$. In view of (iii) in $PM$ $A/orb(k)\in C(orb(k))$ and
in view of (i) also in $V(X)$ $\tilde A/orb(k)\in C(orb(k))$. Hence by (ii)
$\tilde A/orb(k)$ is algebraic, whence $orb(k)$ is finite dimensional.
Therefore in $PM$ by diagonalization $k\in \span EV(A/orb(k))$ and since $k\in
K$ is arbitrary, in $PM$\  $A$ is finitary (and quantum like, since $A$ is
bounded).
\eop
\msk
\msk

\no {2.6. Schr\"odinger equation}.
Our terminology is motivated by the observation, that for
an intrinsically effective
Hamiltonian $T$ the dynamics of the corresponding conservative quantum
system (i.e. $T$ does not depend on the time)
may be evaluated within some permutation model $PM$ of $ZFA$.
Initially we let
the system under observation be prepared in pure state $\sigma(0)\in \dom
T^\circ$.
The state of the system at time $t$, $\sigma (t)$, satisfies
the Schr\"odinger equation
$T\sigma =  i\cdot\partial\sigma/\partial t$. Since
the self adjoint observable $T$ is a generator of the unitary group
$U(t) = exp(-i\cdot T\cdot t)$
in the Hilbert space $K$ of the equation, its solution
is $\sigma(t) = U(t)(\sigma(0))$. As
$T$ is intrinsically effective, up to a unitary equivalence $U:\tilde H\to K$
in $V$ $T = U{\tilde A}U^{-1}$ for some quantum like $A\colon H\to H$
in some permutation model $PM$ such that $U^{-1}\sigma (0) \in
H$ (lemma 16). In $PM$ we set
$V(t)x = exp(-i\cdot\lambda\cdot t)\cdot x$
for $x\in EV(A)$ with $Ax = \lambda x$ and extend $V(t)$ linearly over $H$.
Then
$V(t)\in PM$ is equivalent to the restriction (in fact, a compression)
of the unitary group $U(t)$ to $H$ (lemma 15)
and the solutions of the Schr\"odinger equation
may be evaluated within $PM$ as
$U^{-1}\sigma(t) = U^{-1}U(t)(\sigma(0)) = V(t)(U^{-1}\sigma(0))\in PM$.
\msk

If moreover all eigenspaces of $T$ are finite dimensional, then
independently of $PM$ in $V$ for each bijective isometry $U:\tilde H\to K$
$H = U^{-1}"\dom T^\circ $. In $V$, if $D$ is
any state, then the expected evolution of a commeasurable
quantum system $S = g(T)$ at this state may be computed
within $PM$, too. (In view of lemma 20, if $K$ is separable and $S$ is
intrinsically effective, then $ST = TS$ suffices.)
For there is an operator $B \in PM$ such that
for all Borel functions $f$ the expected value of $f(T)$ at $D$, $tr(f(T)D)$,
is equal to the expected value in $PM$ of $f(A)$ at $B$, $tr(f(A)B)$, where
$tr(C) = \sup \{ \sum _{x\in E}\< Cx,x\>: E$ a finite $ON$--system in $\dom
C \}$. [In $PM$, for $\lambda \in \sigma_p(T)$ we let $P_{\lambda }$ be the
orthogonal projection onto the Cantor complete subspace
$\ker (A - \lambda )$ and $n_{\lambda }$ its
dimension. We set $B = \sum _{\lambda \in \sigma _p(T)}{1\over
{n_{\lambda }}} \cdot tr(U\tilde P_{\lambda }U^{-1}
 D)\cdot P_{\lambda } \in \! PM$;
$tr(P_{\lambda }B) = tr(U\tilde P_{\lambda }U^{-1}D)$.
The extension $U\tilde P_{\lambda }U^{-1}$ of
$P_{\lambda }$ is the projection onto $\ker (T - \lambda )$. Then $tr(f(T)D) =
\sum _{\lambda \in \sigma _p(T)}f(\lambda )tr(U\tilde
P_{\lambda }U^{-1}D) = \sum
_{\lambda \in \sigma _p(T)}f(\lambda )tr(P_{\lambda }B) = tr(f(A)B)$.]
\msk

For example, the Hamiltonian
$T = {\hbar \over 2m}\cdot P^2 + {f \over 2\hbar}\cdot Q^2$
on ${\cal L}_2(\R)$, where $(Pg)(x) = -i\cdot {{dg}\over {dx}} (x)$ and
$(Qg)(x) = x\cdot g(x)$ are the momentum and
position operators, $\hbar  = 1.05\cdot 10^{-27}$ {\it erg sec},
is an unbounded intrinsically effective observable
with one -- dimensional eigenspaces
({\smc Jauch} [Ja68], pp. 211 -- 219, computes an ON--base of
eigenvectors). According to the Ehrenfest rule it corresponds to the
harmonic oscillator of classical mechanics with kinetic energy
${1 \over 2m}\cdot p^2$ and potential ${f \over 2}q^2$.
More generally, by a theorem
due to H. {\smc Weyl}, if in
classical mechanics the potential $v(q)$ is continuous and $\lim v(q) =
\infty$ as $q \to \pm \infty$, then the Hamiltonian of the corresponding
elementary particle is intrinsically effective
(c.f. {\smc Titchmarsh} [Ti62], pp. 110 -- 113, 121 -- 122 and 127)
and all its eigenspaces are finite dimensional ([Du63], p. 1285). Similar
results are true for three dimensional elementary particles.
\msk

The dependence of the representation of the state $D$ by an
operator $B$ in a permutation model $PM$ of $MC$
on the Hamiltonian $T$ is unavoidable even if
$D$ is a mixture of states in $\dom T^\circ$. In $\ell _2(\Z)$ we
let $T$ be an intrinsically effective Hamiltonian with eigenspaces $\span \{
e_k \}$, $k\in \Z $. If $T = U\tilde AU^{-1}$
for some quantum like mapping $A$ on
the Hilbert space $H$ of some model $PM$ of $ZFA+MC$
and some bijective isometry $U:\tilde
H\to \ell _2(\Z )$, then in $V$ $H = U^{-1}"\dom T^\circ =
U^{-1}"\span \{ e_k:k\in \Z \} $ and in $PM$ $H$ is locally
sequentially compact (lemma 8 and $MC$.)
For a sequence $w_k>0$ of weights, $\sum
_{k\in \Z }w_k = 1$, we set $D = \sum_{k\in \Z }w_kP_k$, $P_k$ the orthogonal
projection onto $\span \{e_k +e_{k+1} \} $. Then in $PM$ there is no bounded
symmetric operator $B$ on $H$ such that $tr(U\tilde P U^{-1}D) = tr(PB)$
for all orthogonal projections $P$ of $PM$ in $H$, rather than projections
onto eigenspaces of $T^{\circ}$ only. [For otherwise, when
applied to the projection $P$ onto $\span \{ x\}$, $x\in H$ a unit vector,
$\< DUx,Ux\> = \< Bx,x\>$. The polarization identity
implies $\< DUx,Uy\> = \< Bx,y\> $ for all $x,y\in H$, whence $B$ is the
restriction of $U^{-1}DU$
to $H$. Since $H$ is locally sequentially compact, $H =
\span EV(B) = \span (H\cap EV(U^{-1}DU))$. This contradicts $EV(D)\cap \dom
T^\circ = \emptyset $.] Thus in the notation of quantum theory
lemma 8 proves the impossibility of a certain
type of {\it hidden parameters}.
\msk

As is easily seen from the proof of lemma 18 the position
operator $(Qf)(x) = x\cdot f(x)$ on ${\cal L}_2 ]0,1[$ is far from being
intrinsically effective. [Since by [Ra73], pp. 65 -- 66,
the only finite dimensional invariant subspace
of $Q$ is $\{0\}$, $k\in orb(k) = \{0\}$ for all $k\in K$.] Hence the
classical {\smc Heisenberg} uncertainty relation for the position and the
momentum operators does not apply in an effective quantum theory.
There is, however, a
different phenomenon of complementarity.
\msk

\Lemma  In $ZFA$, there exist unitarily equivalent bounded self adjoint
operators $S, T$ on
$\ell_2(\omega)$ such that $\span EV(S)$ and $\span EV(T)$ are dense, but
$V(S^o,w)\cdot V(T^o,w)\ge {1 \over 4}$ for all unit vectors
$w\in\dom S^o\cap\dom T^o$.
\msk

\Proof. As is wellknown, if one of the self--adjoint observables $S,T$ is
bounded, then
the product of the variances $V(S,w)\cdot V(T,w)$, where
   $$V(S,w)=\<Sw, Sw\> - \<w,Sw\>^2,$$
can be made arbitrarily small at some pure state $w$ (i.e.~$\Vert
w\Vert = 1)$.
\msk

We let $e_i, i\ge 0$, be the standard $ON$ base for
$\ell_2(\omega)$ and $f_i, i\ge 1$, be an $ON$ base for $e_o^\bot$ which
satisfies $\span\{e_i\colon i\ge 1\}\cap \span\{f_i\colon i\ge 1\} = \{0\}$;
$f_o = e_o$. From $e_i$ we define the $ON$ base $w_o = {e_o + e_1
\over \sqrt 2}$, $w_1 = {e_o - e_1 \over \sqrt 2}$, $w_i = e_i$
for $i\ge 2$ and similarily we define $\bar w_i$ from $f_i$. If $S$ is the
diagonal operator $Sw_i = \alpha_iw_i$, all $\alpha_i$ real and
distinct, and $T\bar w_i = \alpha_i\bar w_i$, then $\dom
S^o\cap\dom T^o = \span\{e_o\}$ and $V(S, e_o) = V(T, e_o) = {(\alpha_o -
\alpha_1)^2\over 2}$.
\eop
\msk

Complementarity may be avoided, if the intrinsically effective Hamiltonians
$S$ and $T$ on $K$ are {\it compatible}:
There exist a locally sequentially compact
Hilbert space $H$ in some permutation model $PM$ (c.f. lemma 16),
a surjective isometry $U:\tilde H\to K$ and
quantum like mappings $A,\ B$ on $H$
in $PM$ such that $SU = U\tilde A$ and $TU = U\tilde B$. [If $w$ is an
eigenvector of $A$ or $B$, then $V(S,Uw)\cdot V(T,Uw) = 0$ and $Uw\in \dom
S\cap \dom T$.]
\msk

We recall from spectral theory,
that a concatenation of experiments $S,\ T$ is again an experiment;
i.e. the concatenation corresponds to a Hamiltonian $R$
on $K$ such that (disregarding the
durations of the experiments which we set to be one unit of time) $exp(-
i\cdot S)\cdot exp(-i\cdot T) = exp(-i\cdot R)$. In view of the nonuniqueness
of $R$, an exact specification of the concatenation of noncommeasurable
experiments depends on additional restrictions on the empirical context, such
as compatibility. As an application to control theory we note,
that if $S,\ T$ are intrinsically effective and compatible, then $R$ may be
chosen to be intrinsically effective, too, and compatible with both $S$ and
$T$; i.e. $RU = U\tilde C$ on $\tilde H$
for some bounded quantum like $C\in PM$ on $H$
such that $exp(-i\cdot A)\cdot exp(-i\cdot B) = exp(-i\cdot C)$.
[Since in $PM$ $H$ is locally sequentially compact,
for finite $F\subseteq H$ the submanifold
$E = \span \{ p(A,B)(h):\ p\in \C [x,y],\ h\in F\} $ is
finite dimensional and invariant for both $A$ and $B$.
In $E$ we set ${\cal M} = \{ -i\cdot M:\ M$ a Hermitean matrix$\} =
\{ N:\ N$ skew Hermitean$\}$. By the spectral theorem for normal matrices the
exponential of this Lie algebra is the group $exp''{\cal M} = {\cal U}$ of the
unitary matrices, whence for the restrictions $R = exp(-i\cdot A/E)\cdot
exp(-i\cdot B/E) = exp(-i\cdot C_F)\in {\cal U}$ for some uniquely determined
Hermitean $C_F = f(R)$, where $f\{z:\ \vert z\vert = 1\}\subseteq [0,2\pi [$.
Then by the proof
of lemma 3 in $PM$ $C = \cup \{ C_F:\ F\subseteq H$ finite$\}$ is a symmetric
finitary and everywhere defined bounded
mapping on $H$ which satisfies the desired
exponential equation.]
\msk

As follows from lemma 16, if $T$ and $f(T)$ are self adjoint operators, such
that $T$ is intrinsically effective, then both are extensions of commeasurable
quantum like mappings $A$ and $f(A)$ on some Hilbert space in some model $PM$.
The hypothesis in this observation my be weakened, so that intrinsically
effective observables which are commeasurable in the sense of quantum theory
may be observed simultaneously by means of the measurement of another
intrinsically effective observable and thus are compatible.
\msk

\Lemma {\it In the real world $V$ of $ZFC$, if $S$ and $T$ are intrinsically
effective self adjoint commuting operators on the separable Hilbert space $K$,
then there is a
permutation model $PM$ of $ZFA + MC$ which contains a bounded quantum like
mapping $A$ on a locally sequentially compact Hilbert space $H$, such that in
$V$ $S$ and $T$ are unitarily equivalent (by means of the same isometry)
with functions of $\tilde A$.}
\msk

\Proof. We first observe, that intrinsic effectivity is hereditary. If
$T$ is intrinsically effective and $L$ is a closed invariant subspace, then
the restriction $T/L$ is intrinsically effective on $L$. In view of lemma 16
it suffices to observe, that $\dom (T/L)^{\circ }$ is dense in $L$. For if $P$
is the orthogonal projection onto $L$, $x\in L$ and $x\bot \dom
(T/L)^{\circ}$, then $<x,Py> = 0$ for $y\in EV(T)$ [c.f. the remarks following
lemma 4], whence $<x,y> = <x,Py> + <x,y-Py> = 0$, since $(y-Py)\bot L$;
therefore $x\bot \dom T^{\circ}$ and $x = 0$.
\msk

If $S$ and $T$ are commuting intrinsically effective operators, then for
$\lambda \in \sigma_p (T)$ the closed manifold $L_{\lambda} = \ker (T-\lambda
)$ is a nontrivial invariant subspace for $S$ and we may choose a complete
$ON$--base $B_{\lambda} \subseteq EV(S/L)$
for $L_{\lambda}\subseteq EV(T)$. Since $T$ is intrinsically effective,
$B = \cup \{ B_{\lambda}:\  \lambda \in \sigma_p (T)\}
\subseteq EV(S)\cap EV(T)$ is a complete $ON$--base for $K$ and since $K$ is
separable, it is countable. We let $\beta :B\to \R$ be a bounded injective
mapping and define up to unitary equivalence a bounded
quantum like mapping $A$ on $H = \span B$ in some
model $PM$ as in lemma 16 through $Ab = \beta (b)\cdot b$. Then $f(\beta
(b))\cdot b = Sb$ and $g(\beta (b))\cdot b = Tb$ define real
mappings $f,\  g$ such
that $S,\  T$ are Benioff extensions of $f(A)$ and $g(A)$.
\eop
\msk
\msk

\no {2.7. Quantum logic}.
The {\smc Fraenkel}--{\smc Halpern} model $FH = PM_\Omega$, where $X_\Omega$ is
the algebra of all finite and cofinite subsets of $A$, is known to be the
least permutation model. $FH$ admits {\it amorphous} sets (infinite sets
whose infinite subsets are cofinite), for example, the set $A$ of the atoms is
amorphous, and it satisfies the {\it partial finite
choice axiom} $PAC_{\hbox{\rm fin}}$ (each infinite family of finite sets
admits an infinite subfamily with a choice function.) $FH$ is a source of
several counterexamples.
\msk

In $FH$ the {\it projection lattice} (i.e. the family of the closed
subspaces of $\ell_2 (A)$ which admit orthogonal projections; we do not know
if in general without $AC$ it is a lattice) is {\it modular}
(i.e. it satisfies the shearing identity
$x \wedge (y \vee z) =
x \wedge ((y \wedge (x \vee z)) \vee z)$). {\smc Birkhoff}
and {\smc von Neumann} [Bi36] have considered modularity as a
requirement for quantum logic, but in $ZFC$ it is
satisfied only for systems with a finite degree of freedom.
[For a proof of modularity it suffices to note, that
$S_1 + S_2$ is closed, if the $S_i$ are the ranges of the orthogonal
projections $P_i$. By [Br86], corollary 5.2, when applied to $P_i$,
the subspace $S_i$ is a direct sum of a finite dimensional subspace
and some $\ell_2 (F_i)$, $F_i$ cofinite or empty. Obviously these subspaces
form a lattice. $S_1 + S_2$, as
a sum of finite dimensional spaces and the closed subspace
$\ell_2 (F_1 \cup F_2)$, is closed, whence $S_1+S_2 = S_1\vee S_2$ is the
range of the orthogonal projection $P_1\vee P_2$. Hence for $P_2\le Q$ the
following identity is proved as in [Ha82], solution 15 on p. 177: $(P_1\vee
P_2)\wedge Q = (P_1\wedge Q)\vee P_2$.]
\msk

In $FH$, the Hilbert space $\ell _2(A)$ is a counterexample to the {\smc
Gleason} and {\smc Maeda} [Ma80] theorem about the representation of the
{\it transfinitely additive states} by density matrices.
For if $P$ is a projection, then we set $\alpha (P)
= 0$, if the range of $P$ is finite dimensional, and $\alpha (P) = 1$,
otherwise. In $FH$
for each transfinite sequence $\< P_{\lambda}:{\lambda}\in \kappa \> $
of pairwise orthogonal projections ($P\cdot Q = 0$)
$\alpha (\bigvee _{{\lambda}
\in \kappa }P_{\lambda}) = \sum _{{\lambda}\in \kappa }\alpha (P_{\lambda})
= sup\{ \sum_{\lambda \in K}\alpha (P_{\lambda}) : K\in[\kappa ]^{<\omega}\}$
converges, whence $\alpha $ is a transfinitely additive state. [We
let $e$ be the
least support of the sequence $\< P_{\lambda}:{\lambda}\in \kappa \> $.
In view of [Br86],
corollary 5.2, $P_{\lambda}$ is a direct sum
of a projection in $\ell _2(e)$
and a scalar $\rho _{\lambda}\in \{ 0,1\}$ on $\ell
_2(A\setminus e)$. Hence either all $\rho _{\lambda} = 0$
and $\alpha (\bigvee _{\lambda}P_{\lambda})
= \sum _{\lambda}\alpha (P_{\lambda}) = 0$, or in view of orthogonality exactly
one $\rho _{\lambda} =
1$ and $\alpha (\bigvee _{\lambda}P_{\lambda})
= \sum _{\lambda}\alpha (P_{\lambda}) = 1$. In both cases all except finitely
many projections vanish.] There is, however, no
bounded operator $D$ on $\ell _2(A)$ such that for all orthogonal projections
$P$ the expected value is $\alpha (P) =
tr(PD)$ [If $e$ is the least support of $D$ in $FH$, then $D$ is a direct
sum of a finite matrix on $\ell _2(e)$ and a scalar $\rho $ on $E = \ell
_2(A \setminus e)$. We let $P_E$ be the orthogonal projection onto $E$,
$P_Ex=x\vert E$. Then $\alpha (P_E) = 1$, but $tr(P_ED) = \infty $, if
$\rho \neq 0$, or $tr(P_ED) = 0$, if $\rho = 0$.]
A similar construction shows, that the
{\it ultrafilter extension theorem} $BPI$
does not imply the {\smc Gleason}--{\smc Maeda} theorem for locally
sequentially compact Hilbert spaces, since it fails in the ordered {\smc
Mostowski}--model. [In $\ell_2(A)$, $A$ the set of atoms of that model,
consider $\alpha (P) = 1$, if $\{ a\in A:\! Pe_a=e_a\}$ is unbounded from
above, and $\alpha (P) = 0$, otherwise.]
\msk

The hidden parameter issue involves only the finite
dimensional effective versions of this theorem which, however, are not
constructive. The {\smc Gleason}--{\smc
Maeda} theorem for {\it completely additive states} (additivity for possibly
nonwellorderable families of closed and pairwise orthogonal subspaces)
is effective, too.
\msk

In $FH$ there is a counterexample to
a converse of lemma 6, namely the assertion
{\it "if all quantum like observables on $H$ are algebraic, then
there is an orthonormal base of $H$"} which therefore depends on $AC$.

\Lemma {\it In $FH$ there is a locally sequentially compact Hilbert space $H$
such that each linear map $T: H\to H$,
$\dom T = H$, is a direct sum of a finite
matrix and a scalar (with an infinite dimensional eigenspace).}

\Proof. We set $H = \{x\in\ell_2(A): \sum_{a\in A} x(a) = 0\}$ where $A$ are
the atoms of $FH$. Since by
[Br86], lemmas 2.1 and 2.2 separable subspaces
of $\ell_2 (A)$ are finite dimensional ($A$ is amorphous
and thus $[A]^{<\omega}$ is Dedekind finite; c.f. {\smc Jech} [Je73]), $H$ is
locally sequentially compact and complete. We let $e_a, a\in A$, be the
canonical unit vectors which form an $ON$--base of $\ell_2 (A)$.
\msk

We first observe, that linear functionals on $H$ extend to bounded linear
functionals on $\ell_2 (A)$. In $FH$ there are least supports
([Je73]).
If $f: H\to{\C}$ is any linear functional and $e = supp(f)$ is its least
support, then $f(e_a - e_b) = f(e_a - e_c)$ for $a\in e$, $b,c\in A\backslash
e$. We set $g_a = f(e_a - e_b)$ and $g(x) = \sum_{a\in e} g_a\cdot\<x,
e_a\>$ $(g:\ell_2 (D)\to {\C})$. If $x\in H$, then $f(x) = g(x)$, for if
$x = e_b - e_c$, $b,c\in A\backslash e$, then $f(x) = f(e_c - e_b)$, whence by
linearity
$f(x) = 0 = g(x)$. If $x = e_a - e_b$, $a\in e$ and $b\notin e$, then by the
definition of $g\quad f(x) = g(x)$. If $x = e_a - e_b$, $a\in e$ and $b\in e$,
then for some $c\notin e\quad f(x) = f(e_a - e_c) - f(e_b - e_c) = g(e_a - e_c)
- g(e_b - e_c) = g(x)$ by linearity. Since the vectors $e_a - e_b$ span $H$,
$f = g/H$.
\msk

If $T: H\to H$ is linear and $e = supp(T)$, then $f_a: H\to {\C}$, $f_a(x)
= \<Tx, e_a\>$ is a linear mapping with $supp(f_a)\subseteq e\cup\{a\}$. Hence
the numbers $t(b,a) = \<T(e_b - e_c), e_a\> = f_a(e_b - e_c)$ do not depend on
$c\in A\setminus(e\cup\{a\})$. If we set $\lambda(a) = t(a,a)$ for
$a\in A\setminus e$ and $\lambda(a) = 0$, otherwise, then for $x\in H$ $T$
satisfies
$\< Tx,e_a\> = f_a(x) = \sum_{b\in e} t(b,a)\cdot\<x, e_b\> + \lambda(a)\cdot
\<x, e_a\>$, as follows from the above representation of the linear
functionals. Since the function $t: e\times A\to{\C}$ has support
$supp(t)\subseteq e$, $t(b,a) = 0$ for $a\notin e$.
For $t(b,a) = \< T(e_b - e_c), e_a\>$ for
some $c\notin e$, $c\not= a$, and $y = T(e_b - e_c)\in\ell_2(A)$ satisfies
$y(d) = 0$, if $d\notin supp (y)\subseteq supp T\cup\{b,c\}\subseteq
e\cup\{c\}$;
i.e. $y(a) = \<y, e_a\> = 0$. It follows, that $Tx = \sum_{a\in e}\sum_{b\in
e}t(b,a)\<x, e_b\>\cdot e_a + \sum_{a\in A}\lambda(a)\cdot\<x, e_a\>\cdot e_a$.
Since $\lambda(a) = \lambda$ for some $\lambda\in{\C}$ and all $a\in
A\backslash e$, $T$ is the direct sum of the restriction to $H$ of a finite
matrix $F$ on $\ell_2(e)$, $F(a,b) = t(a,b)$
for $a,b$ in $e$, and the scalar $\lambda$ on
$H\cap\ell_2(A\backslash e)$. As an additional requirement on $F$ we note,
that all column sums are equal to $\lambda$ $(\sum_{a\in A}\<T(e_b - e_c),
e_a\>
= \sum_{a\in e}\<Fe_b, e_a\> - \lambda = 0$, since $T(e_b - e_c)\in H$,
whenever $b\in e,\quad c\in A\backslash e)$.
\msk

There does not exist an $ON$--base $B$ for $H$, for otherwise $H$ is
isomorphic to $\ell_2 (B)$, $[B]^{<\omega}$ Dedekind finite, and
$f(x) = \sum_{b\in B} x(b)$ is an everywhere defined unbounded linear
functional on $\ell_2 (B)$ and thus on $H$.
\eop
\msk

Although $H$ is sequentially closed as a submanifold of $\ell_2 (A)$, it is
dense in $\ell_2 (A)$. [For given $x\in\ell_2(A)$, $\epsilon>0$ and
$E\subseteq A$ finite, such that $\sum_{a\in A\setminus E}\vert
x(a)\vert^2<\epsilon/2$, then we set $y = \sum_{a\in E}x(a)$, choose a set
$F\subseteq A\setminus E$ with $n>{ 2\cdot\vert y\vert\over \epsilon}$
elements and
define $z\in H$ as $z(a) = x(a)$, if $a\in E$; $z(a) = -y/n$, if $a\in F$;
$z(a) = 0$, otherwise. Then $\Vert x-z\Vert_2<\epsilon$ ]. In view of lemma
21, the projection lattice of $H$ is modular and there are counterexamples to
{\smc Gleason}'s theorem. The following construction generalizes lemma 21.
\msk

\Lemma {\it In ZFA, if each locally sequentially compact Hilbert space is
Cantor complete, and $[D]^{<\omega}$ is Dedekind finite, then D is finite.}
\msk

\Proof. Since $[D]^{<\omega}$ is Dedekind finite, we may define
$H = \{x\in\ell_2(D): \sum_{a\in D} x(a) = 0\}$ and $K = H \oplus \ell_2 (D)$.
Since $K$ is locally sequentially compact, it is Cantor complete in view of
our hypothesis. The linear subspace $S=  \{ (x,x):x \in H \}$ is closed,
since it is the graph of the embedding of $H$ into $\ell_2 (D)$, but it does
not admit an orthogonal projection $P$ of $K$ onto $S$. For
if $k=(x,y)\in K$ and $Pk=(z,z)\in S$, $x,\ z\in H,\ y\in
\ell_2 (D)$, then for all $s=(u,u)\in S,\ u\in H$, it holds, that $k-Pk\bot
s$, whence in $\ell_2(D)$ the scalar product
$<x+y-2\cdot z,u>=0$ for all $u\in H$. Hence $x+y+2\cdot z = 0$, which is
impossible for $y\not\in H$. [For if $w\neq 0$, say $w(b)\neq 0$, where $w =
x+y+2\cdot z$, then $<w,u> = w(b)\neq 0$ for the following element $u\in H$:
$u(b)=1$; $u(c)= -1$ for some $c$ such that $w(c)=0$; $u(a)=0$, otherwise.]
{}From the arguments following lemma 4 we conclude, that $K$ is not Cantor
complete, since otherwise each closed linear subspace of $K$ would be the
range of an orthogonal projection in $K$.
\eop
\msk\msk

\no {2.8. Maximal completion}. If $H$ is the Hilbert space of lemma 21,
then $K = FH\cap\tilde H$ is another Hilbert space of $FH$ such that $K\not=
H$. [$H$ is dense in $\ell_2(A)$]. If on the other hand
$H$ is defined in some model $PM$ via lemma 10, then $\tilde H\cap PM = H$. The
same is true for the Hilbert space of {\smc Russell}'s socks. (We note, that
these results depend on the particular {\it canonical embedding} $e\colon
H\to\tilde H$ which is constructed below. Without mentioning $e$ it might be
read as an
abbreviation for: "If in $PM$\  $H$ is dense in the Hilbert space $K$, then $H
= K"$).
\msk

\Lemma {\it If in the permutation model $PM\subseteq V(X)$ $H$ is a Hilbert
space, and if $\tilde H$ is its completion in $V(X)$, then $K = \tilde H\cap
PM\in PM$ and as a submanifold of $\tilde H$ in $PM$ $K$ is a Hilbert space.
If $PM$ satisfies $MC^\omega$, then $K = H$.}
\msk

\Proof. We need to consider a concrete construction of the completion. To this
end in $V(X)$ we let $e\colon H\to\ell_\infty(S_1)$, $S_1 = \{s\in H\colon\Vert
s\Vert = 1\}$, be the embedding $e(h) = \phi_h$, where $\phi_h(s) = <h,s>$.
[Since $\vert\phi_h(s)\vert\le\Vert h\Vert$,
$\Vert\phi_h\Vert_\infty<\infty$.] Then $e$ is easily seen to be a linear
isometry $[\vert\phi_h(s)\vert = \Vert h\Vert$ for $s = h/_{\Vert h\Vert}$ and
$h\not= 0]$. Moreover, since $S_1\in PM$, each function
$\phi\in\ell_\infty(S_1)$ is a subset of $PM$.
\msk

We observe, that ${\cal L} = \ell_\infty(S_1)\cap PM\in PM$. For if $PM$ is
generated by the topological group $G$ and $g\in \hbox{ stab }(H)$,
then for $\phi\in\ell_{\infty}(S_1)$ \par
$\psi = (\hat
d g)(\phi) = \{<(\hat d g)(s), \phi(s)>\colon s\in S_1\}\in V(X)$\par
\nin is a complex
valued function with domain $S_1$ [since $(\hat d g)''S_1 = S_1$] and
$\Vert\psi\Vert_\infty = \Vert\phi\Vert_\infty$. Hence $(\hat d
g)''\ell_\infty(S_1)\subseteq\ell_\infty(S_1)$ and
$\hbox{ stab }({\cal L})\supseteq
\hbox{ stab }(H)$ is open.
${\cal L}$ is $\ell_\infty(S_1)$ in the sense of $PM$ and
the vector space operations and norm restrict from $V(X)$ to the equally
defined functions of $PM$. Moreover, in $PM$ ${\cal L}$ is complete
[completeness of
$\ell_\infty(S_1)$ is provable in $ZFA$.]
\msk

$\tilde H$ is the closure of $e''H$ in the sense of $V(X)$. Since $K = {\tilde
H}\cap PM\subseteq{\cal L}$, in $V(X)$

$$K = \{x\in{\cal L}\colon\forall\epsilon>0\exists h\in H\colon\Vert x-
e(h)\Vert_\infty<\epsilon\}.$$

Similarily, in $PM$ we use this definition to define $K_1\in PM$. Since all
parameters in this definition are in $PM$ and have the same meaning in $PM$ as
in $V(X)$, we conclude, that $K = K_1\in PM$ is the closure of $e''H$ in $PM$.
\msk

The norm and the vector space operations of $K$ are inherited from
$\ell_\infty(S_1)$ and thus from ${\cal L}$, whence in $PM$\  $K$ is a
prae--Hilbert space with
these functions. Since in $PM$\  $K$ is closed in ${\cal
L}$, $K$ is complete.
\msk

Now let $PM$ satisfy $MC^\omega$. In the sequel we shall work inside $PM$. We
pick $x\in K = \tilde H\cap PM\in PM$ and note, that since $e''H$ is dense in
$K$, $W_n = \{h\in H\colon\Vert x - e(h)\Vert_\infty<{ 1\over n}\}$ is nonempty
for $n\ge 1$. By $MC^\omega$ there is a sequence $\emptyset\not= F_n\subseteq
W_n$ of finite sets. We let $y_n$ be the average of $F_n$. Then $y_n\in W_n$,
whence $\lim e(y_n) = x$. Since $H$ is sequentially complete and the sequence
$<y_n\colon n\in\omega>$ is Cauchy [because the sequence $<e(y_n)\colon
n\in\omega>$ is Cauchy], there is some $y\in H$ such that $\lim y_n = y$,
whence $x = \lim e(y_n) = e(y)\in e''H$; i.e. $K = e''H$.
\eop
\msk

Usually the representative of the isometry class of
the completion of $H$, $\tilde H$, will be wellorderable as a set even in
$PM$. Then the embedding
$e : H\to\tilde H$ cannot be traced in $PM$ unless $H$ is
wellorderable and $H = \tilde H$. It is, however, approximated by the
relations $E_G \in PM$; $xE_Gy$, if $x\in H, y\in\tilde H \in V(\emptyset )$
and $y = e(\hat d\pi x)$ for some $\pi\in G$, $G$ an open subgroup of
$\hbox{ stab } H$. $E_G$
induces a partition ${\cal P}_G$ on $\tilde H$; ${\cal P}_G =
\{\tilde H\setminus e" H, \{e(\hat d\pi x) : \pi\in G\} : x\in H\}$.
For example, if in
$FM$ $H$ is the Hilbert space $L$ of {\smc Russell}'s socks, $G =
\hbox{ stab }\emptyset$ and
$\tilde H = \ell_2$, then $xE_Gy$, iff $x(a)\in\{y_n, -
y_n\}$ for $a\in P_n$, $e''H = \{y\in\ell_2 : \{n\in\omega : y_n\not=0\}\in
[\omega ]^{<\omega}\}$ and the ${\cal P}_G$ equivalence class of $y\in e''H$ is
$\{\<\epsilon_ny_n : n\in\omega\> : \epsilon\in\{+1, -1\}^\omega\}$. If it is
assumed, that a pure state in $\tilde H$ provides a complete and exhaustive
description of the system in question, then ${\cal P}_G$ reflects the state of
knowledge about this system due to the information $G$ (e.g. the
orderings on finitely many pairs of socks). The following technical question
is related to these matters. Is $e''H\subseteq\tilde H$ of the first category,
if in $PM$ $H$ is locally sequentially compact? This is true, if $H =
\ell_2(D)$ ({\smc Brunner} [Br86], corollary 2.3.).
\msk

\Lemma {\it If in the permutation model $PM\subseteq V(X)$\  $A$ is a
symmetric linear map whose domain is dense in the Hilbert space $H$ such that
in $V(X)$ there is a unique extension $\tilde A$ of $A$ to a self adjoint
operator on the completion $\tilde H$ of $H$, then $K = \dom{\tilde A}\cap
PM\in PM$ and the mapping $\tilde A/K\colon K\to{\tilde H}\cap PM$ is in
$PM$.}
\msk

\Proof. In $V(X)$, $\tilde A = A^*$ and $\dom A^*$ with the norm $\Vert
x\Vert_1^2 = \Vert x\Vert^2 + \Vert{\tilde A}x\Vert^2$ is a Hilbert space. $H$
is dense in this space, for otherwise the deficiency indices of $A$ could not
vanish ([Du63], p. 1227). The restriction
of the new norm to $H$ is in $PM$ $[\Vert x\Vert^2_1 =
\Vert x\Vert^2 + \Vert Ax\Vert^2$ for $x\in H]$. Hence in $V(X)\ \dom
{\tilde A}$
is the completion of the prae--Hilbert space $(H,\Vert\cdot\Vert_1)$ of $PM$,
whence by the previous lemma $K = \dom{\tilde A}\cap PM\in PM$.
\msk

The mappings $A\colon(H,\Vert\cdot\Vert_1)\to(H,\Vert\cdot\Vert)$ and ${\tilde
A}\colon(\dom{\tilde A}, \Vert\cdot\Vert_1)\to({\tilde H}, \Vert\cdot\Vert)$
are bounded, whence $\tilde A$ is the unique continuous extension of $A$ from
$H$ to $\dom{\tilde A}$. Since in $PM$ this characterization defines ${\tilde
A}/K\colon PM\cap\dom{\tilde A}\to PM\cap{\tilde H}$, $\tilde A/K\in PM$.
\eop
\msk

As an application, we investigate {\it intrinsic intrinsic effectiveness}. If
in $V$ $T$ is a self adjoint operator, such that for some permutation model
$PM$ $T = (T\cap PM)\ \tilde { }$ and in $PM$ $T\cap PM$
is intrinsically effective,
then in $PM$ there is some permutation model $QM$ and a quantum like $A\in
QM$ such that $\tilde A = T\cap PM $. In $V$ there is an $\in
$--isomorphic copy $QM$' of $QM$ which in $V$ is a permutation model
([Br95], theorem 1), whence $T =
(A')\ \tilde { }$ is intrinsically
effective, $A'\in QM'$the quantum like mapping
which corresponds to $A$.
\msk

For Banach spaces, similar investigations about intrinsically
effective operators reveal the noneffective character of the dual spaces,
since the permutation model $PM\subseteq V(S)$
satisfies $AC$, iff the dual $X^*$ in $PM$ of each $B$--space in $PM$ is
dense in the dual in $V(S)$ of the completion $\tilde X$ of $X$.
[If $A\in PM$ is an arbitrary set, then $X = \ell_1(A)$,
formed in $PM$, has the dual $\ell_\infty(A)$ in $PM$, which from the outside
is dense in $B(A,{\cal P}^{PM}(A))$ where $B(A,\Sigma)$ is the space of all
bounded $\Sigma$--measurable complex valued functions on $A$.
Yet $X$ is dense in
$\ell_1(A)$ of
$V(S)$, whose dual is $\ell_\infty(A)$. $B(A,{\cal P}^{PM}(A))$ is dense in
this space, iff ${\cal P}^{PM}(A) = {\cal P}^{V(S)}(A)$; since $A\in PM$ is
arbitrary, $AC$ holds in $PM$]. A conversation by one of the authors with
Prof. {\smc Schachermayer} led to the following related result, whose proof we
have forgotten.
\msk

If ${\cal X}$ is an $\aleph_o$--categorical countable structure over a finite
relational language and $PM\subseteq V(X)$ is generated
by $\hbox{Aut}{\cal X}$ with
the topology of pointwise convergence, then the dual in $V(X)$ of the
completion of the space $\ell_\infty(X)$ of $PM$ is (isometrically) isomorphic
with $\ell_1$, iff ${\cal X}$ is $\omega$--stable.
\msk\msk

\sectionheadline {\bf 3. Appendix}
\msk\msk

\no {3.1. A philosophical problem}.
The reference to permutation models in the definition of intrinsic effectivity
appears to be redundant, since a Hamiltonian
is intrinsically effective, iff it admits a $ON$--base of eigenvectors (proof
of lemma 16.) We conclude this paper with a heuristics which
explains the r\^ole of permutation models.
To this end we shall relate the {\smc Fraenkel--Mostowski} method
to the perception in a wider sense (including the expressive power of the
scientific language, also) of an observer, who wants describe the physical
world. [Thus observables are compatible, if they can be described by the same
observer.] According to {\smc Bohr} [Bo35] {\it ''it lies in the nature
of observation, that all experience must ultimately be expressed in terms of
classical objects."} Hence we may assume, that this world satisfies the laws
of classical physics, thereby avoiding a discussion of "quantum reality".
Despite their speculative character the investigations of section 3 may serve
as a motivation for section 2.
\msk

In order to simplify the discussions about models, we shall require, that if
$T$ is an intrinsically effective Hamiltonian on $K$, then each eigenspace of
$T$ is assumed to be one dimensional. Moreover, $K$ is assumed to be
separable. These requirements are met in most empirically relevant situations.
\msk\msk

\no {3.2. Infinite observers}.
We first investigate
the knowledge which is obtained by means of sensual perception. Following
{\smc Kleene} [Kl56] we identify objects with sets of sensual stimuli. We let
the
set $S$ represent all possible stimuli. Then $U = {\cal P}(S)$ is the set of
all objects (i.e.~things which cannot be distinguished by means of perception
are identified as the same object).
An observer $\omega$ is
described by the set $S_\omega\subseteq S$ of those stimuli which are
perceivable for $\omega$ and the set $U_\omega\subseteq{\cal P}(S_\omega)$ of
objects which are perceived and distinguished by $\omega$. Although
$X_{\omega}$ is finite for each real observer $\omega$, it is plausible to
assume the existence of an infinite set of perceivable objects, too. To this
end infinite {\it coalitions of observers} are
considered which define hypothetical observers $\Omega$
given by $S_\Omega = \cup\{S_\omega\colon \omega\in\Omega\}$ and $U_\Omega =
\cup\{U_\omega\colon\omega\in\Omega\}$. [We apply the theory of section 1.5.
to the infinite set of names $Y = \cup \{\{\omega\}\times X_{\omega}:\
\omega\in\Omega\} $. There $\omega$ corresponds to the
order ideal ${\cal O}_\omega$
for the equivalence relation $R_\omega$ on $U\cup\{a\}$
which describes the state of knowledge of the observer
$\omega\colon$ If $x,y\in U$ are objects such that $x\cap S_\omega = y\cap
S_\omega\in U_\omega$, then $xR_\omega y$, and if $x\cap S_\omega\notin
U_\omega$, then $xR_\omega a$. Now $\Omega$ corresponds to the supremum
${\cal O}_\Omega = \cup\{{\cal O}_\omega\colon\omega\in\Omega\}$.]
\msk

We shall identify $\Omega$ with the set $U_{\Omega}$ of objects and
we shall assume, that $\Omega$ is an infinite Boolean algebra.
If $\Omega$ is the coalition of all possible
observers, then stimuli which do not appear in different objects may be
identified. Thus $\Omega$ is embedded into ${\cal P}(X)$, $X$ the atoms
of the algebra $\Omega$.
\msk\msk

\no {3.3. Meaningful concepts}.
An empirical concept is a higher order
relation involving mathematical constructions and the elements of an
empirical structure ${\cal X}$. In
order to avoid meaningless concepts which in a set theoretical representation
of the concepts might be defined from the $\in$-structure below the elements
of ${\cal X}$, we let these be represented by atoms in the set
theoretical sense. We
define, that an {\it empirical concept} is an element of the
structure $V(X)$, $X$ the base set of ${\cal X}$.
\msk

A relation $R$ on an empirical structure ${\cal X}$ is {\it meaningful} in the
sense
of {\smc Luce} [Lu78], if it is invariant with respect to all automorphisms
of ${\cal X}$. A relation $R$ on ${\cal X}$
is {\it meaningful relative to a finite} set $e\subseteq X$, the base set of
${\cal X}$, if it is invariant with respect to all automorphisms which fix
$e$. If ${\cal X}$ is an $\aleph_\circ$-categorical relational structure, this
is definability from parameters in $e$.
In abstract measurement theory these notions have been extended to
higher order relations, i.e. elements of $V(X)$, as follows.
A group $(G,\cdot)$ is given.
If $d\colon G\to S(X)$ is an injective homomorphism into the
symmetric group over $X$, then $x\in V(X)$ is meaningful, if
$\hat d g(x) = x$ for all automorphisms $g$ of ${\cal X}$.
\msk

A concept is {\it theoretical}, if it is not entirely reducible to perception.
Below we shall relate this vague notion to the precise notion of
meaningfulness. There are, however, empirical structures which admit
meaningless but not theoretical concepts, whence meaningfulness appears to be
too restrictive. [If the empirical structure is a Boolean algebra $\Omega $ of
perceivable objects, then no $x\in \Omega $ is theoretical, but since a
transposition of atoms extends to an automorphism,
$x\in \Omega $  is meaningless, unless $x\in \{ 0,1\} $.]
\msk
\msk

\no {3.4. Evidently theoretical concepts}.
If ${\cal X}$ is an empirical structure which describes the perception, then we
let $PM\subseteq V(X)$ be the permutation model which is generated by $\Aut
({\cal X})$ with the product topology of the discrete topology of $X$ and the
natural group action. This model is related to the problem of what is a
theoretical concept. For we propose, that in the informal definition of a
theoretical concept "reducible to perception" implies "meaningfulness
relatively to finitely many $x\in X$". Moreover, we may assume, that
"entirely" means "for all constituents" of a concept. In the set theoretical
language this translates to "hereditarily" (c.f. {\smc Gandy} [Ga80], where
mechanisms are represented in a similar way).
In view of such assumptions a theoretical concept $x\in
V(X)$ cannot be an element of $PM$.
This finite support model is typical in the following sense. Up to
mutual embeddability of the initial segments
(the $\alpha$-th segment consists of all objects of
rank $\le\alpha$), the class of finite support models coincides
with the class of all permutation models ([Br90], theorems 2.1 and 4.3).
\msk

Thus we are lead to identify an
{\it idealized observer} with a {\smc Fraenkel}--{\smc Mostowski} model
$PM \subseteq V(X)$. The model $PM$ is related to the perception of the
idealized observer $PM$ by means of
the following definition: A concept $x\in V(X)$ is
{\it evidently theoretical} for the observer $PM$, if $x\notin PM$.
In the case of the above finite support model,
this appears to be a sufficient condition
for a concept to be theoretical (intuitively, a theoretical concept is
evidently theoretical, if it is empirically
meaningless for formal reasons already).
In the general case we postulate sufficiency.
In order to relate the abstract model to the real world, we then
need to specify a correspondence rule, for otherwise the condition,
that $x\in V$ is not evidently theoretical does not
represent any restriction, since each concept of the $ZFC$ universe $V$ may be
reconstructed within $V(\emptyset )\subseteq PM$. In celestial mechanics the
notion of evidently theoretical concepts may be traced back to {\smc J.
Kepler}'s "non--entia" in {\it Harmonices mundi libri V} (Linz 1619), where it
is argued, that empirically realizable concepts correspond to constructions by
means of ruler and compass.
\msk

As an
example, let us consider a coalition $\Omega$ whose language of
concept formulation permits for each newly formed
concept the use of finitely many objects only in characterizing its
constituent concepts. As in section 3.2
$X$ is the set of the atoms of $\Omega$ and
$\Omega$ is represented by a set
algebra $X_\Omega\subseteq{\cal P}(A)$, $A$ the atoms of $V(X)$. If the
concept $x\in V(X)$ depends on a finite set $e\subseteq\Omega$ of
objects, then $x$ is meaningful relative to $e$ in the
structure $\Omega$ (i.e. $({\hat d}g)
(x) = x$, if $g\in\Aut(\Omega)$ fixes $e$.) The
concepts $x\in V(X)$ which satisfy this
condition hereditarily are the elements of
the {\smc Weglorz} model $PM_\Omega$.
(In terms of the empirical structure ${\cal X} = \Omega$,
the finite support model of the previous paragraph
is $PM_{\Omega}(\Omega)$ in the notation of [Br95].)
\msk

In view of the proofs of section 2 we need to correlate $FM$ to a specific
type of observation. We consider nerve nets with an infinite past over the
language $L = \{ 0,\ 1\} $. Then a set $W\subseteq L^{<\omega }$ of finite
sequences can be recognized, iff it is definite ({\smc Kleene} [Kl56]:
there is a finite set $E\subseteq \omega $, such that for all words $w_i\in
L^n,\ n\in \omega $, such that $w_0\mid E = w_1\mid E$, $w_0\in W$ and
$w_1\in W$ are equivalent; i.e. $W$ depends on $E$.) We also assume, that
besides the objects (definite sets) the observer perceives the complexity of
the input by means of the relations $xR_ny$ iff $x(n) = y(n)$
and $n\in \dom x = \dom y$. Then the group of
automorphisms of this structure is $\Z^\omega_2$ with the group action
$d$, $(dg)(f)(n) = g(n) + f(n)$ {\it modulo 2} for $n\in \dom f$.
Its topology is
generated by $\hbox{ stab } x = \{g: g/E = 0\}$
where $x$ is a definite set which
depends on $E$. It is the product topology. The corresponding model
$PM$ is closely related to $FM$,
since it is generated by the
same topological group $G$ as $FM$ (but with a different group action),
whence the initial segments of $FM$ can be
$\tilde\in$-isomorphically embedded into $PM$
and vice versa.
\msk\msk

\no {3.5. Correspondence rules}.
We let the Hamiltonian $T$ on $K$ be intrinsically effective. Since $K$ is
separable, in $FM$ there is a Hilbert space $H$ with a quantum like mapping
$A$ on $H$ such that in $V$ $TU = U\tilde A$ for some bijective isometry
$U:\tilde H\to K$ (lemma 16). As is suggested by the preceding discussion,
$FM$ may be identified with an (idealized) observer who investigates the
Schr\"odinger equation $T\sigma = i\cdot {{\partial \sigma }\over {\partial
t}}$. A {\it correspondence rule} between the empirical reality (i.e. quantum
theory on $K$) and its reconstruction in terms of perception (i.e. finitary
quantum theory on $H$) is given by the isometry $U$.
\msk

Modulo this heuristics, a pure state $\sigma (0) \in K$ corresponds to a state
preparation procedure which can be described without a reference to evidently
theoretical concepts, if $U^{-1}\sigma (0)\in \tilde H\cap FM$. As $FM$
satisfies $MC$, $\tilde H\cap FM = H$ (lemma 23), whence $\sigma (0) \in
U"H = \dom T^{\circ} $ (because the eigenspaces of $T$ are finite dimensional)
and $U^{-1}T^{\circ}U\in FM$ (lemma 24).
Therefore both the Hamiltonian $T$ and
the state evolution may be reconstructed within $FM$ (section 2.6),
$U^{-1}\sigma (t) \in H\in FM$, whence the computation of $\sigma (t)$ does
not depend on evidently theoretical concepts. Moreover
if $S$ is intrinsically effective and commeasurable with $T$, then $S = f(T)$
for some Borel function $f$ [since the eigenspaces of $T$ are one
dimensional, this follows from the proof of lemma 20] and the expected value
of the measurement $S$ of $\sigma (t)$ may be determined without the use of
evidently theoretical concepts, even if $\sigma (0)$ depends on such concepts;
$<S\sigma (t),\sigma (t)> = tr(V(-t)f(A)V(t)B)$ for some mixture $B\in FM$
and $V(t) = exp(-i\cdot A\cdot t)$.
(If the eigenspaces of $T$ are finite dimensional but not nececessarily one
dimensional, then the same conclusion
holds with a different correspondence rule.)
\msk

$FM$ therefore supports the intuition, that given a Hamiltonian $T$, a state
preparation procedure which defines a pure state $\sigma (0)$ is a
{\it meaningful input} for $T$ (i.e. the resulting
experimental configuration is not
theoretical), iff $\sigma (0)\in \dom T^{\circ}$. Thus $T$ is intrinsically
effective, iff it is determined by the meaningful inputs [i.e. $\dom
T^{\circ}$ is dense.] In general, however,
this intuition is false. In the model $V(X)$ no state is evidently
theoretical, but $\dom T^{\circ}$ may contain too few states to determine $T$
(lemma 6). We therefore restrict such a heuristics to $FM$ and interpret
intrinsically effective quantum theory as that fragment of constructive
quantum mechanics
which is compatible with the existence of {\smc Russell}'s socks.
\msk\msk

\no {3.6. Conclusion}. The theory of intrinsic effectivity is an
alternative approach to answer questions of constructivity within $ZFC$. It
corresponds to a version of effective mathematics, that replaces the axiom of
choice by the multiple choice axiom.
\msk

The classical approach is the theory of abstract computability of
{\smc PourEl} and {\smc Richards} [Po89] which focuses on the bounded
self adjoint operators on separable Hilbert spaces. Therefore its
applicability as a tool for the working physicist might be restricted
somewhat (but c.f. [Ba93] and [Sc$\infty$]), for
as {\smc Hellman} [He93], p. 240, remarks, unbounded operators are needed
even for the most elementary textbook problems: {\it ``Moreover, typically
such operators are defined on an orthonormal basis (e.g. eigenfunctions of the
Hamiltonian) which at least in standard problems presumably are amenable to a
constructive treatment".} Since such operators (e.g. the harmonic oscillator
which is {\smc Hellman}'s standard example) are intrinsically effective, our
results verify {\smc Hellman}'s ``presumably".
\msk

The Hamiltonians of the hydrogen
atom and the helium atom have continuous spectra and
therefore evade an analysis in terms of intrinsic effectivity. [This concerns
scattering phenomena. The computations in chemistry of energy levels of
electron orbits depend on intrinsically effective direct summands, only.]
Moreover, some fundamental theorems of quantum theory, such as the {\smc
Gleason}--{\smc Maeda} characterization of
the states neither are constructive nor
effective. We therefore again arrive at
{\smc Hellman}'s conclusion (p. 243), that {\it
``the quantum world is indeed a hostile environment for many species of
constructivist mathematics".}
\msk

We conjecture, that this
difficulty might be due to the implications of the constructive analysis of
quantum mechanics for the problem of empirical realizability (c.f. {\smc E.P.
Wigner} in {\it Z. Phys.} 33(1952), p. 101.)
Such speculations, however, transcend the scope of the present paper.
The interpretation of the extra physical notions of our
paper in terms of the perceptions of observers indicates
the r\^ole of intrinsic effectivity in this context.
\msk

Concerning the status of the realizability problem, let us just mention, that
if thought experiments are applied to observables of quantum theory
which are not empirically
realizable, then dubious conclusions may be drawn. An example of incorrect
reasoning is {\smc von Neumann}'s refutation of a hidden parameter theory
also for the spin$1\over 2$ particle (c.f. {\smc Bell}'s rebuttal
[Be66])$^4$.\footnote{}{\hskip-\parindent\vbox{\baselineskip9pt\eightrm\nin$^4$
There are negative results due to
K{\sixrm OCHEN}
and S{\sixrm PECKER} [Ko67], that hidden variable theories do
satisfy a different functional calculus (dependency on parameters encoding the
experimental context) and due to B{\sixrm ELL}, that hidden variable theories
are not local. B{\sixrm OOS} [Bo88]
defines hidden variables for a ground model of set theory in some generic
extension, thereby retaining a strong functional calculus. The theory
of P{\sixrm ITOWSKY} [Pi89] is both almost noncontextual (the lattice
operations are preserved almost surely) and local, but the axioms of
probability theory are weakened.}}
Such concerns over the proper meaning of quantum
mechanics have produced a significant body of
literature, reviewed in {\smc Jammer} [Ja74], which illustrates, that
there is a gap in our knowledge of which mathematical
observables
are empirically realizable.
\bsk
\bsk
\cen{\bftwo References}
\bsk
\frenchspacing

\[Al86  S.~{\smc Albeverio}, J.~E.~{\smc Fenstad},
J.~E.~{\smc Hoegh}--{\smc Krohn} and T.~{\smc Lindstrom}: {\it ``Nonstandard
Methods in Stochastic Analysis and Mathematical Physics"}, Academic Press,
New York (1986).

\[Ba93  M.~{\smc Baaz}, N.~{\smc Brunner} and K.~{\smc Svozil}:
"Interpretations of Combinatory Algebras'', in J.~{\smc Czermak}: {\it
"Philosophy of Mathematics''}, HPT, Wien (1993), 393--406.

\[Be66  J.~S.~{\smc Bell}: "On the Problem of Hidden Variables in Quantum
Mechanics'', {\it Reviews of Modern Physics} 38(1966), 447--452.

\[Be76  P.~A.~{\smc Benioff}: "Models of Zermelo Fraenkel Set Theory as
Carriers for the Mathematics of Physics. I'', {\it Journal of Mathematical
Physics} 17(1976), 618--628.

\[Bo35  N.~{\smc Bohr}: ``Can Quantum--Mechanical Descriptions of Physical
Reality be Considered Complete?", {\it Physical Reviews} 48(1935), 696--702.

\[Bi36  G.~{\smc Birkhoff} and J.~von {\smc Neumann}: ``The Logic of Quantum
Mechanics", {\it Annals of Math.} 37(1936), 823--843.

\[Bo88  W.~{\smc Boos}: Abstract, {\it The Journal of Symbolic Logic} 53
(1988), 1289.

\[Br95  N.~{\smc Brunner}: "A Modal Logic of Consistency'', {\it Rendiconti
Sem. Mat. Univ. Padova} 93(1995).

\[Br90  N.~{\smc Brunner}: "The Fraenkel--Mostowski Method, Revisited",
{\it Notre}\par {\it Dame J. Formal Logic} 31(1990), 64--75.

\[Br86  N.~{\smc Brunner}: "Linear Operators and Dedekind Sets'', {\it Math.
Jap\-onica} 31(1986), 1--16.

\[Br83  N.~{\smc Brunner}: "The Axiom of Choice in Topology", {\it Notre Dame
J. Formal Logic} 24(1983), 305--317.

\[{\smc Br}83 N.~{\smc Brunner}: "Kategories\"atze und multiples
Auswahlaxiom", {\it Zeitschrift f. math. Logik u. G. d. Mathematik (Math. Logic
Quarterly)} 29(1983), 435--443.

\[Br82  N.~{\smc Brunner}: "Dedekind--Endlichkeit und Wohlordenbarkeit",
{\it Monatshefte f. Math.} 94(1982), 9--31.

\[Du57  N.~{\smc Dunford} and J.~T.~{\smc Schwartz}: {\it "Linear Operators
I''}, Interscience Pure and Appl. Math. 7, New York (1957).

\[Du63  N.~{\smc Dunford} and J.~T.~{\smc Schwartz}: {\it "Linear Operators
II''}, Interscience Pure and Appl. Math. 7, New York (1963).

\[Ga80  R.~O.~{\smc Gandy}: "Church's Thesis and Principles for Mechanisms",
in J.~{\smc Barwise} et al.: {\it "The Kleene Symposium"}, North Holland
Studies in Logic 105, Amsterdam (1980), 123--148.

\[Ha82  P.~R.~{\smc Halmos}: {\it "A Hilbert Space Problem Book}," Springer
Graduate Texts in Math. 19, Berlin (1982).

\[He93  G.~{\smc Hellmann}: ``Constructive Mathematics and Quantum Mechanics:
Unbounded Operators and the Spectral Theorem," {\it J. Philosophical Logic
22} (1993), 221--248.

\[Hi84  N.~{\smc Hindman} and P.~{\smc Milnes}: ``The Ideal Structure of
$X^X$", {\it Semigroup Forum} 30(1984), 41--51.

\[Is89 H.~{\smc Ishihara}: "On the Constructive Hahn--Banach theorem", {\it
Bulletin London Math. Soc.} 21(1989), 79--81.

\[Ja74  M.~{\smc Jammer}: {\it ``The Philosophy of Quantum Mechanics}",
Interscience, New York (1974).

\[Ja68  J.~M.~{\smc Jauch}: {\it ``Foundations of Quantum Mechanics"},
Addison--Wesley, Reading (1968).

\[Je73  T.~{\smc Jech}: {\it ``The Axiom of Choice,}" North Holland Studies in
Logic 75, Amsterdam (1973).

\[Ka54  I.~{\smc Kaplansky}: ``{\it Infinite Abelian Groups},"  M.U.P., Ann
Arbor (1954).

\[Kl56  S.~C.~{\smc Kleene}: ``Representation of Events by Nerve Nets", in
C.~E.~{\smc Shannon} et al.: {\it ``Automata Studies"}, P. U. P. Annals of
Math. Studies 34, Princeton (1956), 3--41.

\[Ko67  S.~{\smc Kochen} and E.~P.~{\smc Specker}: "The Problem of Hidden
Variables in Quantum Mechanics'', {\it J. Math. Mech. (Indiana Univ. J.)} 17
(1967), 59--87.

\[Ko65  S.~{\smc Kochen} and E.~P.~{\smc Specker}: "Logical Structures Arising
in Quantum Theory'', in J.~{\smc Addison} et al.: {\it "The Theory of
Models''}, North Holland Studies in Logic 55, Amsterdam (1965).

\[La62 H.~{\smc L\"auchli}: "Auswahlaxiom in der Algebra", {\it Comm. Math.
Helvetii} 37(1962), 1--18.

\[Lu78  J.~D.~{\smc Luce}: ``Dimensionally Invariant Numerical Laws Correspond
to Meaningful Qualitative Relations", {\it Philosophy of Science} 45(1978), 1-
-16.

\[Ma80  S.~{\smc Maeda}: {\it "Lattice Theory and Quantum Logic"}, Makishoten,
Tokyo (1980).

\[Mo91 M.~{\smc Morillon}: "Duaux Continus et Axiome du Choix", {\it Seminaire
d' An\-alyse (Clermont--Ferrand)}, 1991.

\[vN29  J.~{\smc von Neumann}: "Allgemeine Eigenwerttheorie Hermitescher
Funktionaloperatoren'', {\it Mathematische Annalen} 102 (1929), 49--131.

\[Pe93  A.~{\smc Peres}: {\it ``Quantum theory. Concepts and methods"},
Kluwer, Dordrecht (1993).

\[Pe78 A.~{\smc Pelc}: "On some Weak Forms of the Axiom of Choice in Set
Theory", {\it Bull. Acad. Polon. Sc. Ser. Math.} 26(1978), 585--589.

\[Pi89  I.~{\smc Pitowsky}: {\it "Quantum Probability -- Quantum Logic''},
Springer Lecture Notes in Physics 321, Berlin (1989).

\[Pi72 D.~{\smc Pincus}: "Independence of the Prime Ideal Theorem from the
Hahn Banach Theorem", {\it Bulletin American Math. Society} 78(1972), 766--
770.
\[Po89  M.~B.~{\smc PourEl} and I.~{\smc Richards}: ``{\it Computability
in Analysis and}\par{\it Physics"}, Springer Perspectives in Math. Logic,
Berlin (1989).

\[Ra73  H.~{\smc Radjavi} and P.~{\smc Rosenthal}: ``{\it Invariant
Subspaces}", Springer Ergebnisse d. Math. 77, Berlin (1973).

\[Re94 M.~{\smc Reck}, A.~{\smc Zeilinger}, H.~J.~{\smc Bernstein} and B.~{\smc
Bertani}: "Experimental Realization of any Discrete Unitary Operator", {\it
Phys. Rev. Letters} 73(1994), 58--61.

\[Ru85 H.~{\smc Rubin} and J.~E.~{\smc Rubin}: "{\it Equivalents of the Axiom
of Choice, II}", North Holland Studies in Logic 116, Amsterdam (1985).

\[Sa75  G.~{\smc Sageev}: "An Independence Result Concerning the Axiom of
Choice'', {\it Annals Math. Logic} 8 (1976), 1--184.

\[Sc$\infty$ M.~{\smc Schaller} and K.~{\smc Svozil}: "Partition Logics
of Automata", {\it Nuovo Cimento} (to appear).

\[Ti62 E.~C.~{\smc Titchmarsh}:{\it "Eigenfunction Expansions Associated
with Second Order Differential Equations I"}, Oxford U. P., London (1962).

\[Vo79  P.~{\smc Vop\^enka}: {\it ``Mathematics in the Alternative Set
Theory"}, Teubner, Leipzig (1979).

\[Wg69  B.~{\smc Weglorz}: "A Model of Set Theory ${\cal S}$ over a Given
Boolean Algebra'', {\it Bull. Acad. Polon. Sc. Ser. Math.} 17 (1969), 201--
202.

\bye